\documentclass[prb,reprint]{revtex4-2}
\usepackage{amsmath,amssymb}
\usepackage[final]{graphicx}
\usepackage{relsize} 
\usepackage{bm}
\usepackage{overpic}
\usepackage[usenames,dvipsnames]{xcolor}
\usepackage{nicematrix}

\usepackage{siunitx}
\ifdefined\qty\else
  \ifdefined\NewCommandCopy
    \NewCommandCopy\qty\SI
  \else
    \NewDocumentCommand\qty{O{}mm}{\SI[#1]{#2}{#3}}
  \fi
\fi
\ifdefined\unit\else
  \ifdefined\NewCommandCopy
    \NewCommandCopy\unit\si
  \else
    \NewDocumentCommand\unit{O{}m}{\si[#1]{#2}}
  \fi
\fi

\usepackage{natbib}
\usepackage{microtype}

\usepackage[colorlinks,citecolor=red,urlcolor=blue]{hyperref}
\IfFileExists{hypernat.sty}{\usepackage{hypernat}}

\newcommand{\defn}[1]{\emph{#1}}

\newcommand{\bivec}[1]{\tensor{#1}}

\newcommand{\cwbivec}{{\setlength{\fboxsep}{1pt}\fbox{$\circlearrowright$}}}
\newcommand{\ccwbivec}{{\setlength{\fboxsep}{1pt}\fbox{$\circlearrowleft$}}}

\newcommand{\fourvec}[1]{\mathsf{#1}}

\DeclareMathOperator{\tr}{tr}
\newcommand{\dblcdot}{\!:\!}
\DeclareMathOperator{\diag}{diag}

\newcommand{\pvBl}{L} 
\newcommand{\pvBv}{\vec{\pvBl}} 
\newcommand{\bvBl}{\ell} 
\newcommand{\bvBv}{\bivec{\bvBl}} 
\newcommand{\pvCl}{M} 
\newcommand{\pvCv}{\vec{\pvCl}} 
\newcommand{\bvCl}{m} 
\newcommand{\bvCv}{\bivec{\bvCl}} 
\newcommand{\pvDl}{N} 
\newcommand{\pvDv}{\vec{\pvDl}} 
\newcommand{\bvDl}{n} 
\newcommand{\bvDv}{\bivec{\bvDl}} 

\newcommand{\vPl}{U} 
\newcommand{\vPv}{\vec{\vPl}} 
\newcommand{\vQl}{V} 
\newcommand{\vQv}{\vec{\vQl}} 
\newcommand{\vRl}{W} 
\newcommand{\vRv}{\vec{\vRl}} 

\newcommand{\idxl}{a}
\newcommand{\idxm}{b}
\newcommand{\idxn}{c}

\newcommand{\mfrac}[2]{#1/#2}

\begin{document}

\title{Teaching Rotational Physics with Bivectors}

\author{Steuard Jensen}
\email{jensens@alma.edu} 
\author{Jack Poling}
\email{jack.poling99@gmail.com}
\affiliation{Department of Physics and Engineering, Alma College, Alma, MI 48801}

\date{August 17, 2023}

\begin{abstract}
Angular momentum is traditionally taught as a (pseudo)vector quantity, tied closely to the cross product. This approach is familiar to experts but challenging for students, and full of subtleties. Here, we present an alternative pedagogical approach: angular momentum is described using \emph{bivectors}, which can be visualized as ``tiles'' with area and orientation and whose components form an antisymmetric matrix. Although bivectors have historically been studied in specialized contexts like spacetime classification or geometric algebra, they are no more complicated to understand than cross products. The bivector language provides a more fundamental definition for rotational physics, and opens the door to understanding rotations in relativity and in extra dimensions.
\end{abstract}

\maketitle

\section{Introduction}

Rotational physics is full of pitfalls for learners and experts alike. Novice students struggle with right-hand rules and multiple definitions of the cross product, each with non-intuitive features. More advanced students may still stumble over the subtle trap of a left-handed coordinate system or the unexpected behavior of pseudovectors under coordinate reflections. And even experts may not be able to describe angular momentum in relativity or in a theory with extra dimensions of space, since the cross product is unique to three dimensions.

\begin{figure}
\centering
\includegraphics{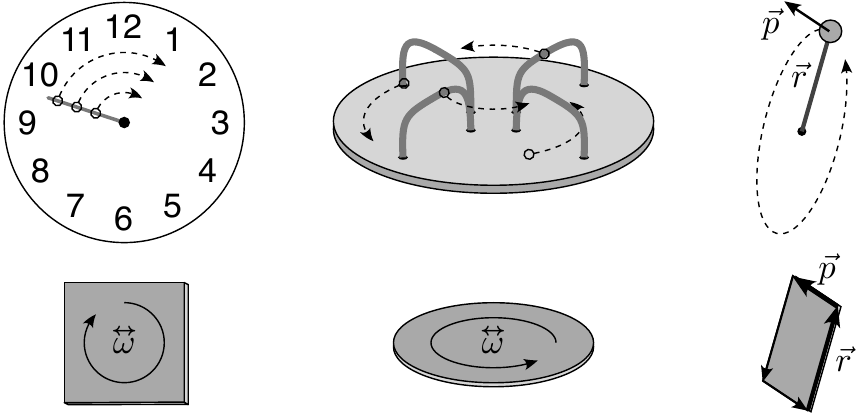}
\caption{\label{fig:tileExamples} Angular velocity $\bivec{\omega}$ is fundamentally a \defn{bivector} quantity, represented as a tile whose attitude and orientation match the object's plane and direction of rotation and whose area gives the magnitude. (The shape of the tile is unimportant.) Here we see angular velocity bivectors for a clock's second hand and a merry-go-round, and the angular momentum bivector (proportional to angular velocity) for a ball on a string. This rightmost tile is constructed as the ``wedge product'' $\vec{r} \wedge \vec{p}$.}
\end{figure}

All of these issues are resolved when we recognize that angular momentum and angular velocity are fundamentally \defn{bivector} quantities.  
Bivectors can be visualized in terms of oriented ``tiles'' whose areas represent their magnitudes, as shown in Fig.~\ref{fig:tileExamples}. Because the tile directly shows the plane and direction of rotation, right hand rules are unnecessary. Coordinate reflections affect the oriented tile in the natural way: there is no extra minus sign as is required for cross products and no need to demand a right-handed coordinate system. And in higher dimensions where there is no concept of an axis of rotation, the idea of a plane of rotation remains meaningful.

Although bivectors are unfamiliar to most physicists, they do appear in the literature. For instance, they play a role in the Petrov classification of spacetimes, \cite{Petrov:1954,Petrov:2000bs} and they are an important component of geometric algebra. \cite{Hestenes:2003,Gull:1993}  In the context of geometric algebra, a text by Doran and Lasenby provides a broad collection of bivector applications.\cite{Doran:2003}  Outside of geometric algebra, describing rotational physics using bivectors (or 2-forms) has mostly occurred in specialized work (general relativity with torsion, for example \cite{Trautman:2006fp}).
Bivectors are, however, applicable to a wide variety of physical systems: any equation involving a pseudovector quantity, a cross product, or a curl can be rewritten in bivector language. 
In addition to rotational systems, bivectors can describe magnetic fields, fluid vorticity, and more.  In fact, the electromagnetic field tensor $F_{\mu \nu}$ is a (four-dimensional) bivector, though it is rarely labeled as one. 

Formally, bivectors are rank-2 antisymmetric tensors (vectors are rank-1 tensors): they are closely related to differential 2-forms (just as vectors are related to 1-forms). But that mathematical description makes them sound more complicated than they are: in practice, understanding bivectors is no more complicated than understanding cross products. Parts of the bivector story may provide students a helpful bridge to familiar cross product results.  Thus, instructors who prefer the traditional treatment may still find bivectors pedagogically useful.

The plan of this paper is as follows. In Section~\ref{sec:Pedagogy} we motivate the reexamination of this fundamental topic and comment on pedagogical advantages and challenges. In Section~\ref{sec:AngMomentumWedge}, we solve a conservation of angular momentum problem using bivectors. In Section~\ref{sec:BivecAddition}, we solve a gyroscopic precession problem by adding bivectors geometrically. In Section~\ref{sec:MatrixProduct} we compute the linear velocity of a point on the Earth's surface by multiplying a vector and a bivector. In Section~\ref{sec:RelativisticAngMomentum} we show how angular momentum in relativity requires a bivector description, and we state general conclusions in Section~\ref{sec:Conclusions}.
In Appendix~\ref{sec:ExtraDims} we solve a conservation of angular momentum problem in extra dimensions.

Additional appendices include
advanced topics and reference material.
Appendix~\ref{sec:Advanced} presents the bivector formalism for the inertia tensor, rotating frames, and quantum commutation relations.
Appendix~\ref{app:Proofs} sketches proofs of two claims in the main text. 
And Appendix~\ref{app:BivecRelations} provides details about products involving pseudovectors and their bivector equivalents.

\section{Why teach bivectors?}
\label{sec:Pedagogy}

To experts with experience studying rotational physics, the bivector description presented below may at first seem unfamiliar and complicated. However, novice learners find the traditional vector approach unfamiliar and complicated as well. 
For instance, two thirds of second-semester physics students tested on cross products answer fewer than half the questions right, regardless of whether the context is electromagnetism or pure math.\cite{Deprez:2019}
Cross products involving vectors pointing out of the plane of the page or right-hand rules requiring awkward arm positions are especially challenging.\cite{Kustusch:2016} (This reference also provides a thorough literature review of educational research on student experiences and difficulties with right-hand rules.)
Furthermore, even after studying rotational motion, the majority of first-semester students still believe that a particle's angular velocity vector points in the plane of motion.\cite{Mashood_2012} 

With these challenges in mind, the bivector formalism offers some clear pedagogical advantages. Representing angular quantities as oriented tiles immediately eliminates the tendency to conflate the vector directions of linear and angular velocity. This representation has a direct, visible relationship to the system's rotation, as opposed to the traditional visualization of an axis of rotation which is more abstract. 
Significantly de-emphasizing the right hand rule removes a major stumbling block for new learners and largely eliminates the need to worry about whether the coordinate system is right handed. 

Moreover, calculating a point particle's bivector angular momentum components as a wedge product (Section~\ref{sec:AngMomentumWedge}) is more intuitive than the cross product procedure. For example, the expression $x p_y - y p_x$ corresponds to the $\ell_{xy}$ component, with the index order matching the positive term (by contrast, the vector component $L_z$ requires students to keep track of cyclic coordinate order). In addition, the tile's area gives a concrete physical meaning to the magnitude formula.%
\footnote{Students accustomed to working with bivector tiles also have an especially easy time understanding the relationship between Kepler's second law and angular momentum, since the triangle formed by $\vec{r}$ and $\vec{v}\,dt$ is essentially just half of the $\protect\bivec{\ell} = \vec{r} \wedge \vec{p}$ parallelogram.} 
The bivector form also generalizes in a straightforward way when studying relativity (Section~\ref{sec:RelativisticAngMomentum}) or extra dimensions (Appendix~\ref{sec:ExtraDims}).

There are four main drawbacks to this approach: two pedagogical, one practical, and one social. First, as Section~\ref{sec:BivecAddition} shows, bivector tile addition is more complicated than tip-to-tail vector arrow addition. In our experience it is hard to justify using the class time necessary to teach this, especially at the introductory level. Second, unlike a vector's components which can be presented as a single column of numbers, a bivector's components are naturally presented as a matrix. This can be daunting since many students have little to no experience with matrices.

In practice, though, both of these difficulties can be largely avoided at the introductory level. The vast majority of rotational and static equilibrium problems lie in a single plane, so only one component of angular momentum or torque is nonzero. In such cases, angular momentum (or torque) is a single signed quantity, and the subtleties of addition in three dimensions don't arise.
The bivector form even has the advantage that the component subscript in (e.g.)\ $\ell_{xy}$ explicitly specifies the positive direction of rotation in the plane.
In those few cases (such as precession) where it is essential to add bivectors in different planes, it may be best to adopt a hybrid approach, where students learn the bivector language but practice converting bivectors to pseudovectors and back when adding them.

In one author's preliminary experience, in order to reduce students' tendency to treat angular quantities as linear velocities or forces it is important for their first encounters with rotations to be based on tiles instead of vectors. 
The practical concern is that all common textbooks and teaching resources use vectors to describe rotational motion, so instructors must develop bivector instructional resources themselves: entire replacements for existing textbook chapters would be best. (A class handout used by the first author includes a table comparing the bivector and vector descriptions, an introduction to the wedge product, and more. It is available on request, and soon as online material hosted by the American Journal of Physics.) 

Attempting to teach the vector and bivector languages simultaneously comes across to students as repetitive and confusing. The most effective approach we have found so far always employs the wedge product (discussed in Section~\ref{sec:AngMomentumWedge}) as the first step in applying the right hand rule: students seem to embrace this approach to cross products. The instructor can decide how much to emphasize that the tiles themselves are bivectors that make the vector form and right hand rule unnecessary.

Probably the most significant concern with adopting the bivector approach is a social one: we are obliged to teach our students how to engage with the larger physics community and the existing literature, which almost always use cross products and pseudovectors. We will address this in full after establishing the bivector formalism. In brief, this change in approach could never happen all at once: there are natural ways for individual instructors to gradually incorporate pieces of the bivector description into their teaching, and this may even help students to better understand the traditional approach.

\medskip

A number of good pedagogical introductions to bivectors already exist, but prior to this work they have all been presented in the context of geometric algebra. 
Examples include articles by Vold,\cite{Vold:1993rot,Vold:1993EM} Hestenes' Oerstead lecture,\cite{Hestenes:2003} and the textbook by Doran and Lasenby.\cite{Doran:2003} 
These sources had no reason to separate the fundamental properties of bivectors from their specific role in geometric algebra, so it is common to see bivectors discussed as if they are \emph{only} defined in that context, as for example in some well-written online introductions.%
\footnote{As of this writing, the Wikipedia entry for bivectors (\url{https://en.wikipedia.org/wiki/Bivector}) is entirely about geometric algebra. Similarly, the website \texttt{\href{https://bivector.net/}{bivector.net}} is actually a geometric algebra site.}

Although one aim of this work is to change that perception and to advocate the value of bivectors within the traditional vector-tensor approach to teaching physics, we also invite geometric algebra teachers to introduce the physical meaning of bivectors in this way before presenting the full geometric product.

\section{Angular momentum and the wedge product}
\label{sec:AngMomentumWedge}

\begin{figure}
\centering
\includegraphics{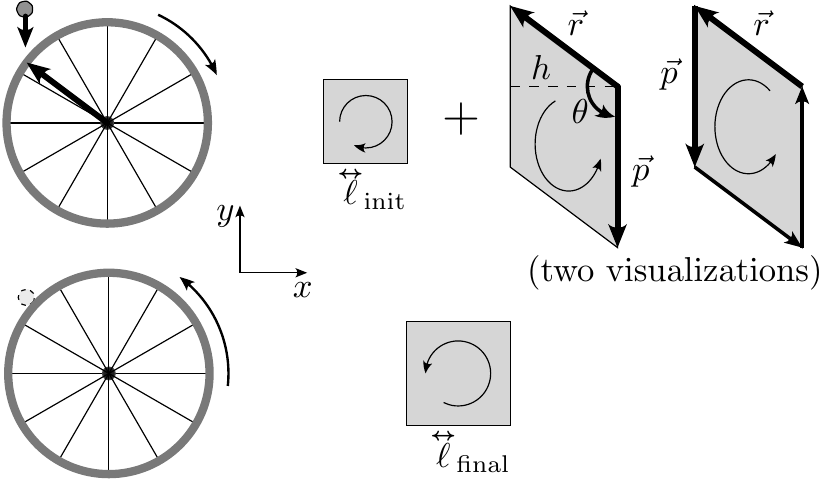}
\caption{\label{fig:WheelWedgeTiles} 
\emph{Top:} Initial configuration: A wheel rotates slowly clockwise as a falling stone is about to hit it. The wheel's initial angular momentum $\bivec{\ell}_\text{init}$ is represented by a small tile. The stone's angular momentum is $\bivec{\ell} = \vec{r} \wedge \vec{p}$: the two vectors define a parallelogram in the plane of the page, of area $|\bivec{\ell}| = h |\vec{p}| = |\vec{r}| |\vec{p}| \sin\theta$. There are two ways to visualize its orientation: either by rotating the first vector toward the second tail-to-tail, or by following the first vector tip-to-tail with the second around the edges.
\emph{Bottom:} Final configuration: The wheel rotates counterclockwise, as the stone falls aside (at rest). The final angular momentum $\bivec{\ell}_\text{final}$ is represented by a tile whose area is the \emph{difference} of the two initial tile areas.}
\end{figure}
To illustrate the bivector formalism in a pedagogical context, consider the angular momentum conservation problem shown in Fig.~\ref{fig:WheelWedgeTiles}. A frictionless wheel of diameter \qty{1}{\meter} and mass \qty{3}{\kilogram} (essentially all at the rim) 
initially rotates clockwise at \qty{1.2}{\radian\per\second}. 
A \qty{0.3}{\kilogram} stone falls straight down and (inelastically) hits the wheel with speed \qty{20}{\meter\per\second} at a horizontal position \qty{0.4}{\meter} left of center. 
To simplify the math, we will assume that the stone comes momentarily to rest 
before falling aside. 
What is the wheel's angular speed when the stone instantaneously comes to rest?

Fig.~\ref{fig:WheelWedgeTiles} illustrates the initial bivector angular momentum of the wheel $\bivec{\ell}_\text{init}$ as an oriented ``tile'' whose \emph{attitude} in space shows the plane of rotation, whose \emph{orientation} (clockwise or counterclockwise) within the plane specifies the direction of rotation, and whose area represents the bivector's \emph{magnitude}. (By definition, the \emph{shape} of the tile is unimportant,%
\footnote{For a bivector defined in terms of the wedge product introduced in this section, it is straightforward to illustrate that changing the parallelogram's shape without changing its area leaves the result unchanged. For example, because the wedge product is linear in each argument, the bivector quantity $6 (\hat{x} \wedge \hat{y})$ can be written as $(6\hat{x}) \wedge \hat{y}$, $(2\hat{x}) \wedge (3\hat{y})$, $(3\hat{x}-\hat{y}) \wedge (2\hat{y})$, or infinitely many other forms, each corresponding to a differently shaped parallelogram with the same area.} 
 and we will see that it can be useful to change from one shape to another.)

\begin{figure}
\centering
\includegraphics{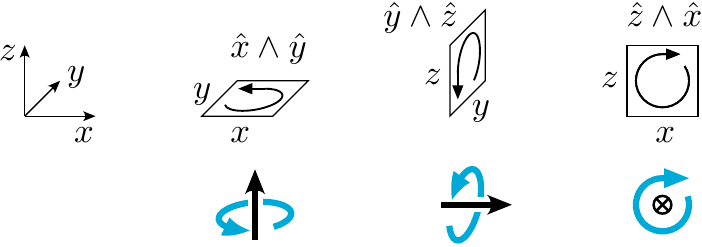}
\caption{\label{fig:CoordBivecs} 
\emph{Top:} Representations of coordinate unit bivectors based on the three dimensional coordinate system shown. \emph{Bottom:} ``Decorated'' vectors representing the same three bivector orientations together with their corresponding normal (pseudo)vectors, related by the right hand rule.}
\end{figure}

Our calculation of the magnitude of $\bivec{\ell}_\text{init}$ proceeds in very much the usual way. 
The moment of inertia of the wheel is $I = M R^2 = \qty{0.75}{\kilogram\meter\squared}$. The initial angular velocity is clockwise in the $xy$-plane: we denote it as $\bivec{\omega}_\text{init} = \qty{1.2}{\radian\per\second}~\cwbivec$, where the symbol $\cwbivec$ indicates a clockwise tile in the plane of the page just as $\otimes$ indicates a vector into the page. (These can be formally interpreted as a ``unit bivector'' and a ``unit vector.'' Unit bivectors in other planes will be illustrated in Fig.~\ref{fig:CoordBivecs}, but there are no convenient text symbols for them.) The initial angular momentum of the wheel is
\begin{equation}
\bivec{\ell}_\text{init} = I \bivec{\omega}_\text{init}
 = \qty{0.9}{\kilogram\meter\squared\per\second}~\cwbivec \quad.
\end{equation}

Modeling the falling stone as a point particle, its (linear) momentum is $\qty{6}{\kilogram\meter\per\second}$ pointing down, and it strikes the wheel at a location \qty{0.5}{\meter} from the axle at an angle $\sin^{-1}(\tfrac{0.4}{0.5}) \approx \ang{53.1}$ from vertical. (Formally, $\vec{p} = -\qty{6}{\kilogram\meter\per\second} \,\hat{y}$ and $\vec{r} = -\qty{0.4}{\meter}\, \hat{x} + \qty{0.3}{\meter}\, \hat{y}$.) We find the bivector angular momentum of the stone as the \defn{wedge product} (or \defn{exterior product}) of position and momentum (rather than the familiar cross product).

The wedge product is illustrated geometrically in two different ways at the top right of Fig.~\ref{fig:WheelWedgeTiles}: the vectors define a parallelogram with a fixed attitude in space, and the order of the vectors determines its orientation (counterclockwise) as described in the caption. The bivector's magnitude corresponds to the parallelogram's area:
\begin{align}
\label{eq:wedge-prod-example}
\bivec{\ell}_\text{stone} &= \vec{r} \wedge \vec{p}
 = |\vec{r}| |\vec{p}| \sin\theta~\ccwbivec \\
 \nonumber
 &= h\, |\vec{p}|~\ccwbivec
 = \qty{2.4}{\kilogram\meter\squared\per\second}~\ccwbivec
\quad,
\end{align}
where $\theta$ is the angle between the two vectors and $h=|\vec{r}| \sin\theta=\qty{0.4}{\meter}$ is the horizontal component of the position.
(This bivector area formula is the underlying reason for the usual cross product magnitude.)
Switching the order of the vectors in the product reverses the orientation while leaving the area and attitude unchanged. We interpret this as reversing the sign of the bivector, so the wedge product is antisymmetric: $\vec{p} \wedge \vec{r} = -\vec{r} \wedge \vec{p}$.

Thus, the system's total angular momentum is
\begin{align}
\nonumber
\bivec{\ell}_\text{total} &= \qty{-0.9}{\kilogram\meter\squared\per\second}~\ccwbivec
 + \qty{2.4}{\kilogram\meter\squared\per\second}~\ccwbivec \\
 &= \qty{1.5}{\kilogram\meter\squared\per\second}~\ccwbivec \quad,
\end{align}
where the first term is negative in the \emph{counter}clockwise direction.
When the stone is instantaneously at rest, this is all carried by the wheel, so its final angular velocity is $\bivec{\omega}_\text{final} = \bivec{\ell}_\text{total}/I = \qty{2}{\radian\per\second}~\ccwbivec$.


Along with this geometric definition, we can also write the wedge product explicitly in components. One natural way to do this is in terms of coordinate unit vectors: for our falling stone,
\begin{align}
\nonumber
\vec{r} \wedge \vec{p}
 &= (-\qty{0.4}{\meter}\, \hat{x} + \qty{0.3}{\meter}\, \hat{y}) \wedge
   (-\qty{6}{\kilogram\meter\per\second} \,\hat{y}) \\
\nonumber
 &= \qty{2.4}{\kilogram\meter\squared\per\second}~\hat{x} \wedge \hat{y}
   - \qty{1.8}{\kilogram\meter\squared\per\second}~\hat{y} \wedge \hat{y} \\
 &= \qty{2.4}{\kilogram\meter\squared\per\second}~\hat{x} \wedge \hat{y}
\quad,
\end{align}
since $\hat{y} \wedge \hat{y} = 0$ by antisymmetry. We can think of $\hat{x} \wedge \hat{y}$ as a ``coordinate unit bivector'' with the orientation that rotates $\hat{x}$ toward $\hat{y}$. This is counterclockwise in the $xy$-plane (seen from above), matching our earlier result.
As shown in Fig.~\ref{fig:CoordBivecs} (top row), these coordinate unit bivectors (labeled with their coordinate edges) can effectively represent bivector orientations on the page.

Using this notation, the full calculation is
\begin{align}
\nonumber
\bivec{\ell}_\text{total} &= \phantom{-{}}
\qty{0.9}{\kilogram\meter\squared\per\second}~\hat{y} \wedge \hat{x} + \qty{2.4}{\kilogram\meter\squared\per\second}~\hat{x} \wedge \hat{y}\\
\nonumber
 &= -\qty{0.9}{\kilogram\meter\squared\per\second}~\hat{x} \wedge \hat{y} + \qty{2.4}{\kilogram\meter\squared\per\second}~\hat{x} \wedge \hat{y}\\
  &= \phantom{-{}}
  \qty{1.5}{\kilogram\meter\squared\per\second}~\hat{x} \wedge \hat{y}
  \quad,
\end{align}
(where the wheel's initial motion is in the orientation that rotates $\hat{y}$ toward $\hat{x}$).

We can also label components with subscripts: $\bivec{\ell}_\text{total} = \ell_{xy} \, \hat{x}\wedge\hat{y}$, with $\ell_{xy} = \qty{1.5}{\kilogram\meter\squared\per\second}$. For any wedge product such as $\bivec{\ell} = \vec{r} \wedge \vec{p}$,
\begin{equation}
\label{eq:bivecIndices}
\ell_{ij} = r_i p_j - r_j p_i \quad,
\end{equation}
where the indices $i$ and $j$ run over the coordinates $x,y,z$. (Note that this implies the shorthand $r_x = x$, $r_y = y$, $r_z = z$.) The components are antisymmetric: $\ell_{ji} = -\ell_{ij}$.

Intuitively, the reason that Eq.~\eqref{eq:bivecIndices} requires two terms on the right is that $\bivec{\ell}$ can receive contributions to (e.g.)\ $\ell_{xy}$ (in the $xy$ plane) in two ways: when $\vec{r}$ extends in the $x$ direction and $\vec{p}$ extends in the $y$ direction, and also (with the opposite orientation) when $\vec{r}$ extends in the $y$ direction and $\vec{p}$ extends in the $x$ direction. 

These components form an antisymmetric matrix:
\begin{align}
\nonumber
\bivec{\ell} &= 
\begin{pmatrix}
 0 & \ell_{xy} & \ell_{xz} \\
 \ell_{yx} & 0  & \ell_{yz} \\
 \ell_{zx} & \ell_{zy} & 0
\end{pmatrix}
= \begin{pmatrix}
 0 & \ell_{xy} & -\ell_{zx} \\
-\ell_{xy} & 0  & \ell_{yz} \\
\ell_{zx} & -\ell_{yz} & 0
\end{pmatrix} \\
\label{eq:wedgeComponentIndices}
&= \begin{pmatrix}
 0 & x p_y - y p_x & x p_z - z p_x \\
y p_x - x p_y & 0  & y p_z - z p_y \\
z p_x - x p_z & z p_y - y p_z & 0
\end{pmatrix} \\
\nonumber
&= \begin{pmatrix}
 0 & \qty{2.4}{\kilogram\meter\squared\per\second} & 0 \\
-\qty{2.4}{\kilogram\meter\squared\per\second} & 0  & 0 \\
0 & 0 & 0
\end{pmatrix}
\quad,
\end{align}
where the values in the last line are specific to the stone in our example. In the second matrix of the first line we have used antisymmetry to write all of the components in terms of the three independent $\ell_{ij}$ whose indices are in cyclic order.

It is worth emphasizing that students who are unfamiliar with matrices can easily just work with individual component equations exactly as one can with vectors; the indices show exactly which components to multiply and in what order. 




\begin{figure}
\centering
\includegraphics{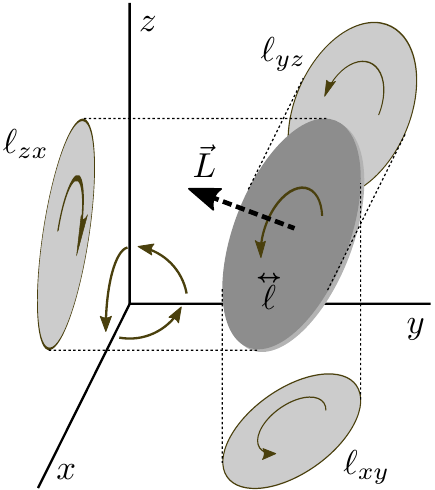}
\caption{\label{fig:tileProjections} The tile representing $\bivec{\ell}$ is projected 
onto each of the three coordinate planes. 
Comparing the orientation of each projection to the plane's orientation (shown here in cyclic coordinate order), in this example we can see that $\ell_{xy}$ and $\ell_{yz}$ are positive while $\ell_{zx}$ is negative. (If we instead considered the other orientation for the $xz$-plane, we would say $\ell_{xz}$ was positive.)
Also shown is the normal vector $\vec{L}$ to the tile, which is equivalent to the usual cross product: its direction relates to the orientation of $\bivec{\ell}$ by the right hand rule.}
\end{figure}

The bivector components $\ell_{ij}$ have a geometric meaning: they are the projections of the tile's area onto each corresponding coordinate plane, as in Fig.~\ref{fig:tileProjections}. The order of indices specifies an orientation of the coordinate plane: the one that rotates the first coordinate into the second. Then the sign of each component indicates whether the projected orientation matches that orientation. Because the coordinate order is explicit, the component formulas in Eq.~\eqref{eq:bivecIndices} are valid whether or not the coordinate system is right-handed.

Because the Pythagorean theorem applies to coordinate plane area projections of a flat tile just as it does to lengths,\cite{Conant:1974gp} the magnitude of a bivector is
\begin{equation}
\label{eq:bivecMagnitude}
|\bivec{\ell}| = \ell_{xy}^2 + \ell_{yz}^2 + \ell_{zx}^2
 = \frac{1}{2} \sum_{i,j} \ell_{ij}^2
 \quad,
\end{equation}
where the factor of $\frac{1}{2}$ corrects for double counting when we sum over all possible index pairs (in both orders). 
(This is the ``double dot product'' of the bivector with itself: see Eq.~\protect\eqref{eq:pseudo-dot-pseudo-doubledot}.)

Although it is not a \emph{necessary} part of studying bivectors, Fig.~\ref{fig:tileProjections} also illustrates that (in three dimensions) every bivector tile has a unique normal vector whose ``length'' equals the tile's ``area'' and whose direction can be found from a right hand rule by curling your fingers in the sense of the tile's orientation. 
This relationship provides us with a particularly clear way of indicating bivector orientations on the page: as shown in Fig.~\ref{fig:CoordBivecs} (bottom row), the bivector orientation can be drawn as a curved arrow \emph{around} its normal vector.
(These ``decorated'' vectors are sometimes used in engineering to represent torque.)

The components of this normal vector match specific bivector components:
\begin{equation}
\label{eq:crossProdComponents}
\vec{L} = \begin{pmatrix} \ell_{yz} \\ \ell_{zx} \\ \ell_{xy} \end{pmatrix}
 = \begin{pmatrix} y p_z - z p_y \\ z p_x - x p_z \\ x p_y - y p_x \end{pmatrix}
 \quad.
\end{equation}
The first equality applies to any bivector with known components. For bivectors constructed as a wedge product $\vec{r} \wedge \vec{p}$, the second equality also applies and this can serve as the \emph{definition} of the cross product $\vec{r} \times \vec{p}$. The bivector component labels in this definition provide helpful reminders of the sometimes baffling cross product component formulas.
In this context, a right handed coordinate system becomes necessary, but of all the right hand rules in physics, this one 
may be the simplest to understand, and it is explicitly connected to the intuitive rotation direction of the bivector tile.

\begin{figure}
\centering
\includegraphics{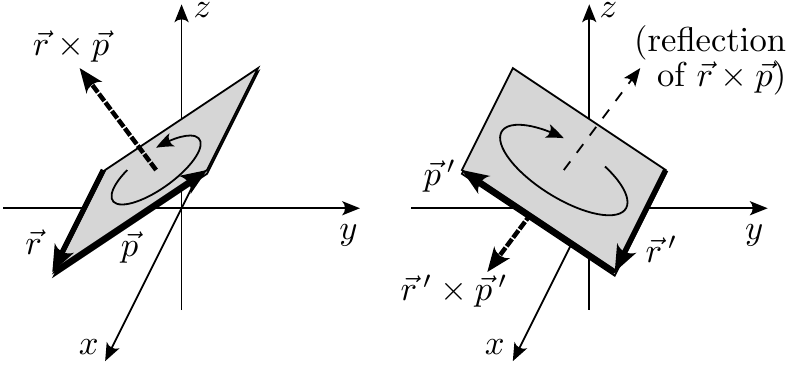}
\caption{\label{fig:bivecReflection} \textit{Left:} The bivector tile $\vec{r} \wedge \vec{p}$ together with the cross product $\vec{r} \times \vec{p}$. \textit{Right:} The same vectors after reflection $y \to -y$, now labeled $\vec{r}\,'$ and $\vec{p}\,'$. The bivector tile $\vec{r}\,' \wedge \vec{p}\,'$ and its orientation have reflected across the $xz$-plane in the natural way. But the cross product $\vec{r}\,' \times \vec{p}\,'$ is \emph{not} simply the reflection of $\vec{r} \times \vec{p}$ (shown lightly dashed): instead, its direction must also be reversed to agree with the right hand rule. This is the defining behavior of a \defn{pseudovector}.}
\end{figure}

Formally, the quantity $\vec{L}$ constructed from a bivector in this way is a \defn{pseudovector} (or \defn{axial vector}), as distinguished from normal (or \defn{polar}) vectors. It behaves like a vector under rotations, but under a reflection like $y \to -y$ its overall direction reverses (in addition to the reflection) as shown in Fig.~\ref{fig:bivecReflection}: $(L_x, L_y, L_z) \to (-L_x, L_y, -L_z)$. 
This awkward behavior of pseudovectors becomes entirely natural in bivector components, where the components whose signs change are precisely the ones involving $y$: $(\ell_{yz}, \ell_{zx}, \ell_{xy}) \to (-\ell_{yz}, \ell_{zx}, -\ell_{xy})$. This corresponds to reversing the signs of the $y$-row and $y$-column of $\bivec{\ell}$ (following the standard tensor transformation law).

\section{Precession and bivector addition}
\label{sec:BivecAddition}

It is more complicated to apply the above methods in a fully three dimensional context. 
\begin{figure}[t]
\centering
\includegraphics{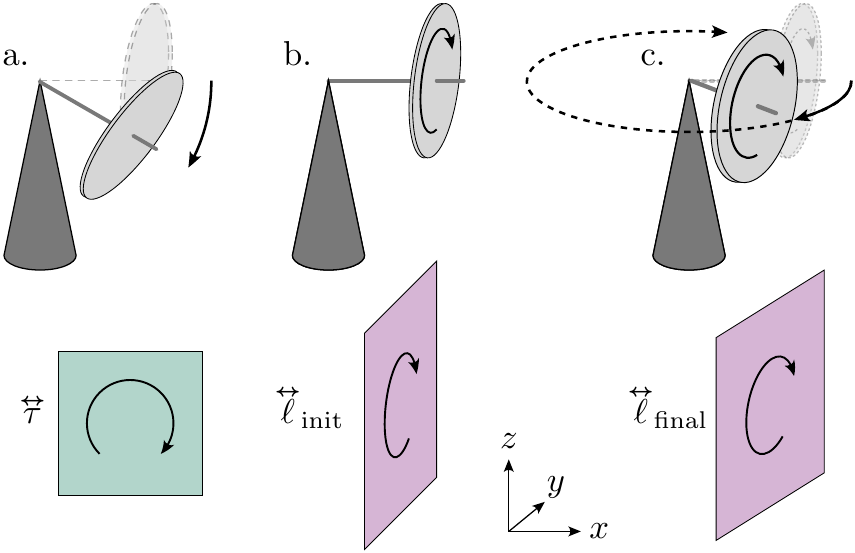}
\caption{\label{fig:precessionSetup} Gyroscopic precession. (a) shows the torque due to gravity as a tile $|\bivec{\tau}|\, \hat{z}\wedge\hat{x}$: the plane of rotation if there were no initial angular momentum. (b) shows the initial angular momentum, $|\bivec{\ell}_\text{init}| \,\hat{z} \wedge \hat{y}$. (c) shows the gyroscope precessing in the horizontal plane: details are given in Sec. \ref{sec:BivecAddition}.}
\end{figure}
As a specific example, consider the situation illustrated in Fig.~\ref{fig:precessionSetup} of a gyroscope disk with mass 0.2kg, and moment of inertia 0.001 ${\rm kg\, m^2}$, and which rotates around its axis at 117 rad/s.  It is supported by a horizontal massless rod with its center \qty{12}{\centi\meter} from the frictionless pivot. What is its precession frequency?

As shown in the figure, the initial torque due to gravity about the pivot point can be visualized as a tile oriented with the natural plane of rotation if the gyroscope were not spinning and dropped from rest: a bivector with orientation $\hat{z}\wedge\hat{x}$. In particular, since the force of gravity is $\vec{F} = m \vec{g} = -\qty{1.96}{\newton}\, \hat{z}$ acting at position $\vec{r} = \qty{0.12}{\meter}\, \hat{x}$, 
the torque is
$\bivec{\tau} = \vec{r} \wedge \vec{F} = \qty{0.2352}{\newton\meter}~\hat{z}\wedge\hat{x}$.
Meanwhile, the gyroscope's initial angular momentum tile matches its plane of rotation: 
$\bivec{\ell}_\text{init} = I\bivec{\omega} = \qty{0.117}{\kilogram\meter\squared\per\second}~\hat{z}\wedge\hat{y}$.

Torque is defined as $\bivec{\tau} \equiv d\bivec{\ell}/dt$. 
We will use the relationship $d\bivec{\ell} = \bivec{\tau} dt$ to see how the gyroscope's angular momentum changes from $\bivec{\ell}_\text{init}$ to $\bivec{\ell}_\text{final} = \bivec{\ell}_\text{init} + d\bivec{\ell}$ in time $dt$ and will find the angle $d\phi$ between the two planes.

%

\begin{figure*}
\centering
\includegraphics{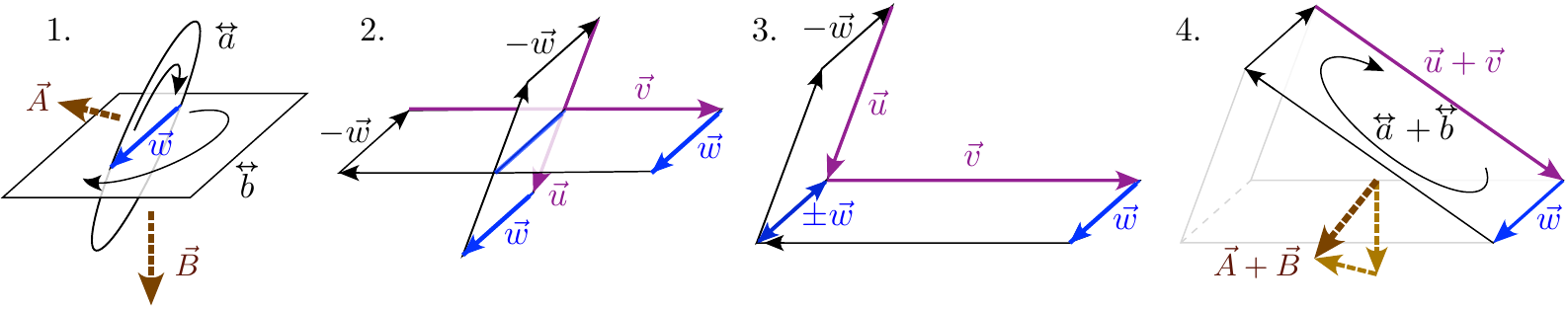}
\caption{\label{fig:tileAdd} The four-step process for adding two tiles to represent the sum of bivectors $\bivec{a}$ and $\bivec{b}$: 1.~\emph{Line of intersection}. 2.~\emph{Reshape to rectangles} ($\vec{u}\wedge\vec{w}$ and $\vec{v}\wedge\vec{w}$): areas and orientations are unchanged. 3.~\emph{Shared edge cancels}: overlap $+\vec{w}$ and $-\vec{w}$ edges, so $\vec{u}$ and $\vec{v}$ are tip-to-tail. 4.~\emph{Close the triangular prism:} the new face is the result, $(\vec{u}+\vec{v})\wedge\vec{w}$. (If you imagine the tiles of step~3 as soap bubbles, the $\pm\vec{w}$ edge has now been removed and the bubble has snapped tight.) 
The simpler pseudovector approach to bivector addition is shown in step~1 and step~4, dashed arrows represent the pseudovectors corresponding to each bivector tile.}
\end{figure*}

It is possible to solve the problem purely algebraically, but to actually understand what is going on we need to understand bivector addition geometrically. The simplest way to do this is to convert to traditional vector addition. One can find each bivector tile's pseudovector equivalent, add those vectors tip-to-tail, and then convert the result back to a tile. (The new tile is normal to the result of the sum and has area equal to its magnitude; its orientation is determined by the right hand rule.) This is illustrated alongside the first and last steps in Fig.~\ref{fig:tileAdd}: the resulting argument would be identical to traditional vector-based discussions of precession.

If, however, we want to work exclusively in the bivector language, there is also a natural geometric interpretation of addition for bivector tiles themselves. 
If the tiles are parallel, one can use the procedure described in Section~\ref{sec:AngMomentumWedge}. Otherwise, the general process for adding arbitrary bivectors $\bivec{a}$ and $\bivec{b}$ as tiles is illustrated in Fig.~\ref{fig:tileAdd}, and proceeds as follows:
\begin{enumerate}
\item \textbf{Line of intersection:} Let the tiles overlap, and find the line where the two tiles intersect. Formally, we can choose a vector $\vec{w}$ along this line.
\item \textbf{Reshape to rectangles:} Reshape each tile into a rectangle with the line of intersection ($\vec{w}$) as one edge. The other edges (denoted $\vec{u}$ and $\vec{v}$) should have appropriate length and directions to keep the area and orientation the same. This is equivalent to choosing $\vec{u}$ and $\vec{v}$ so that $\bivec{a} = \vec{u} \wedge \vec{w}$ and $\bivec{b} = \vec{v} \wedge \vec{w}$.

\item \textbf{Shared edge cancels:} Line up the two tiles so the shared edges overlap \emph{with opposite orientation}(the two edges ``cancel out''). If you follow the orientation of each tile around its boundary, $\vec{u}$ and $\vec{v}$ should be tip-to-tail. 

\item \textbf{Close the triangular prism:} Replace the ``bent'' shape with a new flat tile that closes the triangular prism formed by the earlier rectangles. Algebraically, this is the sum $\vec{u} \wedge \vec{w} + \vec{v} \wedge \vec{w} = (\vec{u}+\vec{v}) \wedge \vec{w}$.

\end{enumerate}
(See Appendix~\ref{app:FormalTileAdd} for a formal discussion of how to identify appropriate vectors $\vec{w}$, $\vec{u}$, and $\vec{v}$.)
There's no denying it: this process of adding tiles is subtle, and for addition it's usually easier to just work in components or with the pseudovector equivalents. But, it is still good to know that there \emph{is} a meaningful geometrical interpretation.

\begin{figure}[t]
\centering
\includegraphics{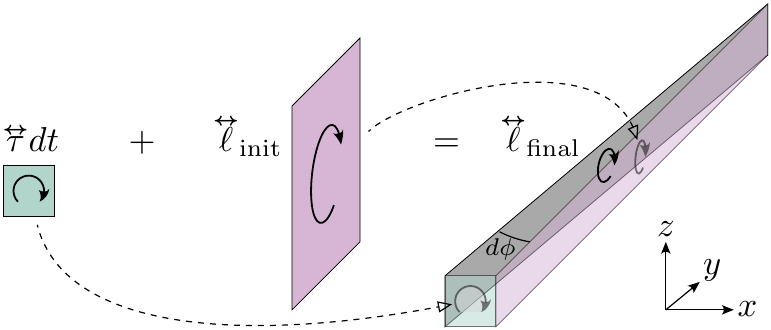}
\caption{\label{fig:precessionTilesOnly} Analyzing gyroscopic precession from Fig.~\ref{fig:precessionSetup} as a sum of orthogonal tiles. 
The torque is multiplied by a small time $dt$, resulting in an area much smaller than the initial angular momentum. Adding the two requires reshaping the $\bivec{\ell}_\text{init}$ tile: we can treat the shared vertical edge as $\hat{z}$, so the length of each horizontal edge is the tile's magnitude.
The resulting final angular momentum is then $\bivec{\ell}_\text{final} = |\bivec{\tau}| dt\, \hat{z}\wedge\hat{x} + |\bivec{\ell}_\text{init}|\, \hat{z}\wedge\hat{y} = \hat{z}\wedge \bigl(|\bivec{\ell}_\text{init}|\,\hat{y} + |\bivec{\tau}| dt\, \hat{x} \bigr)$.
Its orientation has rotated horizontally by angle $d\phi$ as the gyroscope's plane of rotation precesses, corresponding to the top edge's rotation from $|\bivec{\ell}_\text{init}|\, \hat{y}$ to $\bigl(|\bivec{\ell}_\text{init}|\,\hat{y} + |\bivec{\tau}| dt\, \hat{x} \bigr)$.}
\end{figure}

Once this tile addition process is understood, we benefit by being able to directly visualize what it means to ``add'' two different planes of rotation (without the extra conceptual step of changing to the perspective of axes of rotation and back). With this, we can solve our precession problem. Fig.~\ref{fig:precessionTilesOnly} shows the torque and angular momentum tiles that were introduced in Fig.~\ref{fig:precessionSetup}, with the torque multiplied by a short time $dt$.
In the final step, the tiles have been reshaped to have the same height (corresponding to the shared vector $\vec{w}$ discussed earlier: we choose this to be $\hat{z}$), so each bivector's magnitude is proportional to its tile's length in the $xy$-plane. Thus, the angle $d\phi = \tan^{-1}( |\bivec{\tau}| dt/|\bivec{\ell}_\text{init}|) \approx |\bivec{\tau}| dt/|\bivec{\ell}_\text{init}|$ (where we have used the small angle approximation since $dt$ is very short), giving a precession frequency of
$\omega_P \equiv \frac{d\phi}{dt} = |\bivec{\tau}|/|\bivec{\ell}_\text{init}| \approx \qty{2.0}{\radian\per\second}$.

For a student who has become comfortable with adding tiles, this explanation of precession may be conceptually easier to grasp than the usual explanation based on axes of rotation. It can often feel indirect and complicated to translate the visible rotational motion and the torque into vectors, then formally add those vectors, and finally translate the result back to a new visible plane of rotation. But the process of adding tiles is undeniably complicated as well: we are trading one set of challenges for another.

\section{Rigid rotations and the matrix product}
\label{sec:MatrixProduct}

\begin{figure}
\centering
\includegraphics{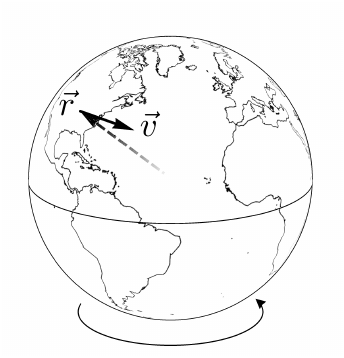}
\hspace{-4mm}
\includegraphics{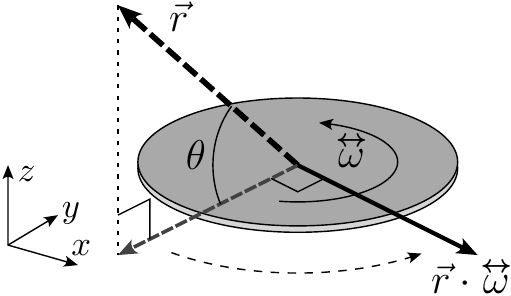}
\caption{\label{fig:EarthVecTileMult} \emph{Left:} The vector $\vec{r}$ points from the center of the earth to the city of Chicago's position. The city's instantaneous linear velocity is $\vec{v}$: parallel to the plane of the equator going east.
\emph{Right:} The process for finding the (dot) product of a vector $\vec{r}$ with a bivector $\bivec{\omega}$. First, project the vector into the plane of the bivector, and then rotate it \ang{90} in the tile's orientation direction. The resulting magnitude is $|\vec{r}|\,|\bivec{\omega}|\,\cos\theta$. (This matches the formula for cross product magnitude, $|\vec{\omega}\times\vec{r}|=|\vec{\omega}|\,|\vec{r}|\,\sin(\ang{90} - \theta)$.)}
\end{figure}

Rigid body rotations allow us to demonstrate one more core part of bivector math. Consider the following question, illustrated in Fig.~\ref{fig:EarthVecTileMult}: The city of Chicago is located \ang{42} north of the equator. Taking the Earth to be a sphere of radius \qty{6400}{\kilo\meter}, what is Chicago's linear velocity relative to the center of the Earth? (This problem can of course be solved using elementary methods, but we will use it to illustrate the last essential bivector operation.)

Introductory discussions of rigid rotating objects generally include the relationship between the linear and angular speeds of a point on the object: in bivector language, $|\vec{v}| = r |\bivec{\omega}|$ for a point a distance $r$ from the axis of rotation. This is the magnitude part of a vector equation traditionally written as a cross product: $\vec{v} = \vec{\omega} \times \vec{r}$.  Translating this equation into bivector language cannot result in a wedge product because we start with a bivector and end with a vector. In components, $\bivec{\omega}$ is a square matrix, while $\vec{v}$ is a vector.

Instead, we express this relationship as a \emph{dot} product (matrix product) of the vector and the bivector to yield a vector: 
\begin{equation}
\label{eq:rotationVelocityBivec}
\vec{v} = \vec{\omega} \times \vec{r}
 \quad \longrightarrow \quad
\vec{v} = \vec{r} \cdot \bivec{\omega}
\quad.
\end{equation}
The order of terms is opposite that in the traditional cross product equation, and order does matter because of the antisymmetry of the bivector, $\vec{r} \cdot \bivec{\omega} = -\bivec{\omega} \cdot \vec{r}$.

The geometric interpretation of the product $\vec{r} \cdot \bivec{\omega}$ consists of two steps, as shown 
in Fig.~\ref{fig:EarthVecTileMult}.
First, project the vector $\vec{r}$ into the plane of the bivector. In our case, the result is a vector of magnitude $|\vec{r}| \cos\theta = \qty{6400}{\kilo\meter} \cos\ang{42} \approx \qty{4756}{\kilo\meter}$ pointing horizontally away from the Earth's axis of rotation. (The factor of $\cos\theta$ comes from the projection, exactly as in the vector dot product.) Second, multiply that vector by $|\bivec{\omega}|$ and rotate the result \ang{90} in the direction of the bivector's orientation (so the result is perpendicular to $\vec{r}$).%
\footnote{Intuitively, the idea behind this projection-rotation process is that the bivector is an area spanned by one vector sweeping around to a second vector. By taking a dot product with an additional vector from the left, we replace the first vector with a scalar, rotating the result toward the second. This can be made formal by reshaping the tile as a wedge product of a unit vector along the additional vector's projection and a second vector perpendicular to it. (A dot product from the right would rotate the resulting vector \ang{90} \emph{opposite} the bivector's orientation, reversing the sign.)} 
In our case, $|\bivec{\omega}| = 2\pi\,\unit{\radian}/\qty{24}{\hour} \approx \qty{7.27e-5}{\radian\per\second}$. 
Then the result has magnitude $|\vec{v}|=|\vec{r}|\,|\bivec{\omega}|\,\cos\theta \approx \qty{346}{\meter\per\second}$, and the direction is due east. 

In component language, we interpret $\vec{r} \cdot \bivec{\omega}$ as the matrix product of a row vector and a square matrix. Because it is more familiar to multiply a square matrix times a column vector, we will often reverse the order with a minus sign: $\vec{v} = -\bivec{\omega} \cdot \vec{r}\,$.%
\footnote{Related to this, a rotation matrix can be constructed as the matrix exponential $\mathsf{R}(t) = e^{-\protect\bivec{\omega} t}$. Then for a point with initial position $\vec{r}_0$, the matrix product $\vec{r}(t) = \mathsf{R}(t) \cdot \vec{r}_0$ gives the rotated position as a function of time.} 
Then in general,
\begin{align}
\label{eq:MatrixProdComponents}
\begin{pmatrix} v_x \\ v_y \\ v_z \end{pmatrix}
 &= -\begin{pmatrix}
 0 & \omega_{xy} & \omega_{xz} \\
\omega_{yx} & 0  & \omega_{yz} \\
\omega_{zx} & \omega_{zy} & 0
\end{pmatrix}
\begin{pmatrix} x \\ y \\ z \end{pmatrix} \\
\nonumber
 &= -\begin{pmatrix} 
   \omega_{xy} y + \omega_{xz} z \\ 
   \omega_{yx} x + \omega_{yz} z \\ 
   \omega_{zx} x + \omega_{zy} y
 \end{pmatrix}
 = \begin{pmatrix} 
   -\omega_{xy} y + \omega_{zx} z \\ 
   \omega_{xy} x - \omega_{yz} z \\ 
   -\omega_{zx} x + \omega_{yz} y
 \end{pmatrix}
\;.
\end{align}
For our particular case Earth's rotation is in the $\hat{x}\wedge\hat{y}$ orientation; choosing the prime meridian as the $+x$ direction, Chicago`s position is 
$\vec{r} \approx \left(166, -4750, 4280\right)\unit{\kilo\meter}$, 
since its longitude is \ang{88}~W. Thus
\begin{align}
\begin{pmatrix} v_x \\ v_y \\ v_z \end{pmatrix}
 &\approx -\begin{pmatrix}
 0 & 7.27 & 0 \\
-7.27 & 0  & 0 \\
0 & 0 & 0
\end{pmatrix}
\begin{pmatrix} \phantom{-{}}166 \\ -4750 \\ \phantom{-{}}4280 \end{pmatrix}
\times 10^{-5} \,\unit{\kilo\meter\per\second}
\;,
\end{align}
with the result 
$\vec{v} = \left(346, 12, 0\right)\unit{\meter\per\second}$: 
east from Chicago.

All of these results match the traditional cross product answers. 

\section{Relativistic angular momentum}
\label{sec:RelativisticAngMomentum}

In special relativity, vector quantities are understood as the spatial parts of corresponding four-vectors in $(3+1)$ dimensions. Position, $\vec{r},$ is combined with time as $\fourvec{X} = (ct, x, y, z)$, and momentum $\vec{p}$ is the spatial part of the four-momentum, whose time component is the energy: $\fourvec{P} = (E/c, p_x, p_y, p_z) = \gamma m \,(c, v_x, v_y, v_z)$. 

However, simply adding an initial ``timelike'' component to make a four-vector doesn't work for pseudovectors: the cross product is undefined in four dimensions, and the relativistic transformation law for $\vec{L}$ is entirely different than for vectors like $\vec{p}$ or $\vec{r}$. 
The relativistic angular momentum of a point particle must be expressed as a four-bivector $\bivec{\fourvec{M}} = \fourvec{X} \wedge \fourvec{P}$, with (contravariant) components~\cite{Penrose:2004rr}
\begin{align}
\nonumber
\bivec{\fourvec{M}} &= 
\begin{pmatrix}
 0 & -c N_x & -c N_y & -c N_z \\
 c N_x & 0 & \ell_{xy} & -\ell_{zx} \\
 c N_y & -\ell_{xy} & 0  & \ell_{yz} \\
 c N_z & \ell_{zx} & -\ell_{yz} & 0
\end{pmatrix}
\quad.
\end{align}
The spatial components are the angular momentum bivector $\bivec{\ell}$, and the time-space components form a (true) vector $\vec{N}$ in three dimensions. 
The standard Lorentz transformations of this rank-2 tensor lead to exactly the correct transformation laws for angular momentum, which mix the components of $\bivec{\ell}$ with those of $\vec{N}$.

To understand $\vec{N}$, we can write out the wedge product definition: $N_i = x_i \frac{E}{c^2} - t p_i = \gamma m (x_i - v_i t)$. When summed over a collection of particles, this means that the vector $\vec{N}$ is effectively total energy times ``the apparent location of the center of mass at $t=0$.'' (We say ``apparent'' location because the extrapolation in time to $t=0$ treats the current center of mass velocity as a constant.)

Unfamiliar as this is, in the absence of external forces the center of mass velocity doesn't change, so this ``$t=0$ position'' is a conserved quantity, just like the (spatial) angular momentum. It may feel odd to see a specific time ($t=0$) as a defining aspect of a conserved quantity, but we have measured angular momentum about $\vec{r}=0$ as well: all components of relativistic angular momentum depend on our choice of spacetime origin.


Having defined relativistic angular momentum it is natural to ask whether there is also a bivector form for relativistic angular velocity, but it turns out not to have a self-consistent relativistic definition. Just as idealized rigid objects are not consistent with relativity, whether or not an object is undergoing rigid rotation is not a relativistically invariant question.%
\footnote{One way of understanding this is that velocities in relativity do not add in a linear way: the Einstein velocity transformation means that different atoms of a rotating object will have their velocities changed by different amounts when boosted to a different frame, so Eq.~\protect\eqref{eq:rotationVelocityBivec} will no longer be true (even relative to the center of mass).} 


\section{Conclusions}
\label{sec:Conclusions}

It is more than a little audacious to propose a radical change in how we think about and teach a fundamental topic like rotational motion.  The existing formalism is effective and familiar. There will always be a need to teach students the cross product and the pseudovector formalism: even if the bivector approach presented here were universally adopted, students would still encounter cross products throughout the literature. It is easy to argue that for those very real social reasons, such a change would be both impractical and unnecessary.

Nevertheless, it is the perspective of this work that cross products in physics always hide some underlying bivector phenomenon, and that there is value in exposing that truth to our students. The bivector description of rotational physics is approachable enough to be taught at a wide range of levels, and even experts may gain fresh insight into this familiar topic.
(The same may also be true for the bivector description of the magnetic field%
.\cite{Jensen:2023mag})

Despite those benefits, a substantial change in approach doesn't happen all at once. Rather than completely rewriting lesson plans and textbooks, we could begin by planting seeds of the bivector description during traditional cross product instruction. That might mean showing students how to construct an oriented tile as an intermediate step to finding the cross product vector, both to justify its magnitude and to offer an easier right hand rule. 
It might mean drawing torque and angular momentum on the board as ``decorated'' vectors (see Fig.~\ref{fig:CoordBivecs}) as a reminder that they are not quite like ordinary vectors.
It might mean setting up a problem in the $xy$-plane and labeling the torque ``$\tau_{xy}$'' rather than ``$\tau_z$'' or ``$\tau$''. Or it might mean commenting on the existence of the bivector description and some of its advantages during an upper level class, in hopes that the students will recognize the concept if they encounter it again.

We have provided instructional resources which provide students explicit practice with non-planar bivector manipulations: the precession calculation in Section~\ref{sec:BivecAddition} is one example of this; an even simpler problem might be a collision between a point particle and an already-rotating sphere. (The example in Appendix~\ref{sec:ExtraDims} could easily be reduced to three dimensions for this purpose.)

Describing rotational quantities as bivectors can be helpful to students while also providing a deeper conceptual understanding. It is our hope that this perspective will eventually become one small part of how the field as a whole thinks about rotational physics.

\begin{acknowledgments}
Thanks to  Brian Hancock and May Lee for very helpful comments on earlier drafts of this work, which have led to significant improvements in clarity and accessibility. 
The image of Earth in Fig.~\ref{fig:EarthVecTileMult} was made with Mathematica.
\end{acknowledgments}

\section*{Author Declaration}
The authors have no conflicts to disclose.


\appendix

\section{Rotations in extra dimensions}
\label{sec:ExtraDims}

String theory and other ideas in modern physics invite us to study physics in systems with more than three dimensions of space. While this topic is not a part of the traditional undergraduate curriculum, it is a fun and fascinating idea that many students are eager to hear about, and for the most part it requires only simple extensions of the familiar principles of mechanics. As we will see, the bivector formalism is unavoidable when studying rotational motion in this context.

Most of the basic laws of physics extend in a straightforward way: for example, if $d=4$ then the momentum vector would have components $\vec{p} = (p_x, p_y, p_z, p_w)$. But just as with spacetime in relativity, the pseudovector description of rotational physics is not even well-defined in that case: there is no concept of a unique axis of rotation. 

In $d$ dimensions of space there are $\binom{d}{2} = \frac{d(d-1)}{2}$ independent planes of rotation: the number of distinct ways to choose 2 out of $d$ coordinates. Each of these has its own bivector component. For example, when $d=4$ we can write the angular velocity as
\begin{equation}
\label{eq:bivecComponents4D}
\bivec{\omega} = 
 \begin{pmatrix}
0 & \omega_{xy} & \omega_{xz} & \omega_{xw} \\
-\omega_{xy} & 0 & \omega_{yz} & \omega_{yw} \\
-\omega_{xz} & -\omega_{yz} & 0 & \omega_{zw} \\
-\omega_{xw} & -\omega_{yw} & -\omega_{zw} & 0
\end{pmatrix} \quad.
\end{equation}
Because there are more independent components than the number of dimensions, there is no way to represent angular velocity as a vector. (This is why $d=3$ is the unique special case where a cross product can be defined.)

The greater number of planes allows rotations to be substantially more complicated than what we are used to in three dimensions. One aspect of this is that only rare bivectors in four or more dimensions can be represented as a single tile. The addition rules in section~\ref{sec:BivecAddition} give a single tile result for the case of two parallel tiles (such as $\hat{x}\wedge\hat{y} + 3\hat{x}\wedge\hat{y}$) and for two tiles that overlap along a line (such as $\hat{x}\wedge\hat{y} + \hat{z}\wedge\hat{x}$, which overlap along a line parallel to the $x$-axis and yield the bivector $\hat{x}\wedge(\hat{y}-\hat{z})$). But most tiles in four dimensions are entirely separate and overlap only at a single point (such as $\hat{x}\wedge\hat{y} + \hat{z}\wedge\hat{w}$) 
and those sums cannot be simplified at all.
(A \defn{simple bivector} is one that can be written in the form $\vec{u}\wedge\vec{v}$; this is sometimes called a \defn{blade}.)

In general, any bivector in $d$ dimensions can be written as a sum of $d/2$ vector wedge products (rounded down), each represented by a tile. So in three dimensions it is always possible to represent a bivector as a single tile, but in four or five dimensions most require a sum of two tiles, and so on. 
It is even possible to require that all of the vectors in those wedge products be mutually orthogonal (so that each tile is orthogonal to all the others). Unless two tiles have the same magnitude, that orthogonal sum is unique. (A sketch of a proof is given in Appendix~\ref{app:OrthogDecomposition}.)

A concrete example of angular momentum conservation may help clarify all this. %
Consider a (hyper)spherical rock of mass \qty{3.7e4}{\kilogram} and moment of inertia \qty{1}{\kilogram\meter\squared} floating in space with its initial angular velocity (about the center of mass) in a single plane: 
$\bivec{\omega}_\text{init} = \hat{y} \wedge (10 \hat{x} + 16 \hat{z}) \,
  \unit{\radian\per\second}$ (about three rotations per second).
A pebble of mass \qty{1}{\kilogram} with velocity
$\vec{v} = (0,2,-1,0)\,\unit{\kilo\meter\per\second}$
collides with the rock at position
$\vec{r} = (4,0,8,-1)\,\unit{\milli\meter}$ relative to the rock's center of mass.
We want to know the rock's final angular velocity once the pebble is embedded in it.

The rock's initial angular momentum is just $I \bivec{\omega}_\text{init}$: the numerical components match those of its angular velocity.
The pebble's angular momentum is the wedge product $\vec{r}\wedge\vec{p}$, where $\vec{p} = m\vec{v} = (0,2000,-1000,0)\,\unit{\kilogram\meter\per\second}$. Then
\begin{align}
\vec{r}\wedge\vec{p} &=
\left(\!\begin{smallmatrix}
0 & 8& -4 & 0 \\
-8 & 0 & -16 & 2 \\
4 & 16 & 0 & -1 \\
0 & -2 & 1 & 0
\end{smallmatrix}\!\right) \unit{\kilogram\meter\squared\per\second}
\quad.
\end{align}
This means that the final total angular momentum is
\begin{align}
\nonumber
\bivec{\ell}_\text{final} &=
\left[
\left(\!\begin{smallmatrix}
0 & -10 & 0 & 0 \\
10 & 0 & 16 & 0 \\
0 & -16 & 0 & 0 \\
0 & 0 & 0 & 0
\end{smallmatrix}\!\right) 
+
\left(\!\begin{smallmatrix}
0 & 8& -4 & 0 \\
-8 & 0 & -16 & 2 \\
4 & 16 & 0 & -1 \\
0 & -2 & 1 & 0
\end{smallmatrix}\!\right) 
\right] \unit{\kilogram\meter\squared\per\second} \\
 &=
\left(\!\begin{smallmatrix}
0 & -2 & -4 & 0 \\
2 & 0 & 0 & 2 \\
4 & 0 & 0 & -1 \\
0 & -2 & 1 & 0
\end{smallmatrix}\!\right) \unit{\kilogram\meter\squared\per\second}
\quad.
\end{align}
After dividing by the moment of inertia (which is essentially unchanged), we can write the final angular velocity as $\bivec{\omega}_\text{final} =
  \bigl[ -2\hat{x}\wedge(\hat{y}+2\hat{z}) + (2\hat{y}-\hat{z})\wedge\hat{w} \bigr] 
  \unit{\radian\per\second}$. This can be visualized as a sum of two rectangular tiles that are entirely orthogonal to each other: a simultaneous rotation in two different planes, one twice as fast as the other.%
\footnote{If we do not demand that the two planes be mutually orthogonal, there are many other ways to write this final angular velocity as a sum of two wedge products. For example, $\bigl[ (2\hat{x}+\hat{z})\wedge(\hat{y}+2\hat{z}) + (2\hat{y}-\hat{z})\wedge(\hat{y}-\hat{w} ) \bigr] \unit{\radian\per\second}$.}

The math of bivectors has interesting implications for the nature of rotations in four (or more) dimensions. In four dimensions, rotations can be categorized as simple, double, or isoclinic. Simple rotations have angular velocities given by simple bivectors: they are rotations parallel to a single plane, just as in three dimensions. Double rotations, by far the most common, have angular velocities that can only be written as the sum of two tiles: at any point but the origin, all four coordinates are changing at once under simultaneous rotations in the two tile planes. And isoclinic rotations are the special case of double rotations where the two tiles have the same angular speed: in this case, there are infinitely many ways of dividing the motion into two orthogonal planes of rotation.

One application of this is solar system formation, when a cloud of randomly moving dust collapses to form a star and structures in orbit around it. In three dimensions every bivector is simple, so the total angular momentum of any system lies in a single plane. Collisions between particles will eventually cancel out most motion orthogonal to that plane, concentrating the matter into a dense rotating disk where planets can form. In four dimensions angular momentum is usually not a simple bivector, so there is no preferred plane where the matter will collect and it will remain in a diffuse cloud where planet formation is unlikely.\cite{minutephysics:flat} (This difficulty is in addition to the radial instability of circular orbits in higher dimensions.)

\section{More advanced applications}
\label{sec:Advanced}

The bivector approach to rotational mechanics described above continues to be viable for more sophisticated topics, and can in fact make some relationships look arguably more natural. Here, we will sketch just a few of those applications as results without proof: in most cases, they can be deduced from the traditional forms using the methods of Appendix~\ref{app:BivecRelations}.

\subsection{The inertia tensor}
\label{sec:InertiaTensor}

In the traditional 
formulation, the general relationship between angular velocity $\vec{\omega}$ and angular momentum $\vec{L}$ is given by a symmetric rank-2 tensor $I_{ij}$ (which can be interpreted as a linear map from vectors to vectors). In index notation, this relationship is $L_i = \sum_j I_{ij} \omega_j$. The equivalent relationship in bivector language is
\begin{equation}
\label{eq:bivector-L-omega}
\ell_{ij} = \frac{1}{2} \sum_{k,l} I_{ijkl} \, \omega_{kl} 
\quad.
\end{equation}
The factor of $\frac{1}{2}$ is the usual correction for double-counting when summing over entries of an antisymmetric matrix, just as in Eq.~\eqref{eq:bivecMagnitude}. 
The rank-4 inertia tensor $I_{ijkl}$ can be interpreted as a linear map from bivectors to bivectors. 

The bivector formalism leads directly to a formula for the components $I_{ijkl}$. Combining Eqs.~\eqref{eq:wedge-prod-example} for $\bivec{\ell}$ and~\eqref{eq:rotationVelocityBivec} for rigid rotations, a particle of mass $m$ in a rigid body with angular velocity $\bivec{\omega}$ has angular momentum $\bivec{\ell} = \vec{r} \wedge (m\, \vec{r} \cdot \bivec{\omega})$. Summing over the entire rigid body, we find
\begin{equation}
\bivec{\ell} = \int dm\, \vec{r} \wedge (\vec{r} \cdot \bivec{\omega}) \quad.
\end{equation}
We can convert this to index notation and then use the Kronecker delta to factor out $\bivec{\omega}$:
\begin{align}
\nonumber
\ell_{ij} &= \int dm \left( \sum_{k} \left( x_i (x_k \omega_{kj}) -  x_j (x_k \omega_{ki}) \right) \right) \\
 &= \sum_{k,l} \left( \int dm \left( x_i x_k \delta_{jl} - x_j x_k \delta_{il} \right)\right) \omega_{kl}
 \quad,
\end{align}
where we have reversed the order of the sum and integral in the last step. Comparing this to Eq.~\eqref{eq:bivector-L-omega}, the term in large parentheses is one half times the inertia tensor. As our last step, because the bivector $\omega_{kl}$ is antisymmetric in $k$ and $l$, we will also insist that the inertia tensor have the same antisymmetry. (Any symmetric component would cancel out in the sum). The final result is:%
\footnote{Index notation provides an efficient shorthand for this result: $I_{ijkl} = 4 \intop dm \, x_{[j} \delta_{i][k} x_{l]}$, where square brackets denote antisymmetrization: $M_{[ij]}=\frac{1}{2} (M_{ij} - M_{ji})$.}
\begin{align}
  \nonumber
I_{ijkl} &= \int dm \, \left( x_i x_k \delta_{jl} - x_j x_k \delta_{il}
  - x_i x_l \delta_{jk} + x_j x_l \delta_{ik} \right) \\
\label{eq:bivector-inertia-tensor}
 &= \int dm \, \bigl( x_i x_k \delta_{jl} + (\text{symmetries}) \bigr) 
\quad.
\end{align}
In this final form, all four terms are essentially the same: they just use the bivector antisymmetry to  sum over the different ways that an $\ell_{ij}$ index could match an $\omega_{kl}$ index.
Thus, for example, $I_{xyxy} = \int dm \, (x^2 + y^2)$ and $I_{yzzx} = \int dm\, (-yx)$.
It is straightforward to show that this tensor has some simplifying symmetries:
\begin{gather}
I_{ijkl} = -I_{jikl} = -I_{ijlk} = I_{klij} \\
  I_{ijkl} + I_{iklj} + I_{iljk} = 0
\;.
\end{gather}
The antisymmetry of the initial and final pairs matches that of the bivectors, and the symmetry under exchange of pairs corresponds to the symmetry of the inertia tensor in the usual formulation.\footnote{Curiously, these are exactly the algebraic symmetries of the Riemann curvature tensor.}

In this language, rather than finding eigenvectors of $I_{ij}$ to identify principal axes of rotation, we find \defn{eigenbivectors} of $I_{ijkl}$ to identify principal \emph{planes} of rotation. In brief, as in the Petrov classification of spacetimes,\cite{Petrov:1954,Petrov:2000bs} to do this we treat the space of bivectors as an abstract vector space and create a matrix $I_{AB}$ showing the action of $I_{ijkl}$ on the basis bivectors $\hat{y}\wedge\hat{z}$, etc. (In $d$ dimensions of space, as discussed in Appendix~\ref{sec:ExtraDims}, these bivector space indices range from $A=1,\ldots,d(d-1)/2$.) In three dimensions, one natural way to do this is to map bivector index pairs to pseudovector indices so that $I_{AB}$ becomes just the traditional inertia tensor $I_{ij}$: e.g.\ $I_{xy} = I_{yzzx}$.

To illustrate this process, consider a uniform solid cube with one corner at the origin and sides of length $a$ oriented along the positive coordinate axes. We can compute the components of its inertia tensor for rotations about the origin, using the density $\rho=M/a^3$ to define $dm = \rho \,dx\,dy\,dz$. For example,
\begin{align}
I_{xyxy} &= \int_0^a \!dx \int_0^a \!dy \int_0^a \!dz \left( x^2 + y^2 \right)
  = \frac{2}{3} M a^2  \\
I_{xyxz} &= \int_0^a \!dx \int_0^a \!dy \int_0^a \!dz \left( yz \right)
  = \frac{1}{4} M a^2 \quad.
\end{align}
Because of the cube's symmetry, up to a $\pm$ sign all other nonzero components will equal one of these two results. 

To find the principle planes of rotation, we define a basis $\bivec{b}_A$ for the space of bivectors (with $A=1,2,3$) by%
\footnote{This ``alphabetical order'' basis is one natural choice. If instead we chose the ``cyclic'' basis $(\hat{y}\wedge\hat{z}, \hat{z}\wedge\hat{x}, \hat{x}\wedge\hat{y})$, the corresponding pseudovector basis would be the traditional $(\hat{x}, \hat{y}, \hat{z})$ and the components of $I_{AB}$ would be the same as in a traditional calculation.}
\begin{equation}
\bivec{b}_A = (\hat{x}\wedge\hat{y},\hat{x}\wedge\hat{z},\hat{y}\wedge\hat{z}) \quad,
\end{equation}
so a general bivector can be written $\bivec{\omega} = \sum_{A=1}^3 \omega_A \bivec{b}_A$. We define the linear map on this space as
\begin{equation}
I_{AB} \equiv \frac{1}{4} \sum_{i,j,k,l} (b_A)_{ij} \, I_{ijkl} \, (b_B)_{kl} \quad,
\end{equation}
where the factor of $\frac{1}{4}$ compensates for double counting when summing over two antisymmetric pairs of indices. For our system, putting this together with the results above gives
\begin{equation}
I_{AB} = M a^2 \begin{pmatrix}
\mfrac{2}{3} & \mfrac{1}{4} & -\mfrac{1}{4} \\
\mfrac{1}{4} & \mfrac{2}{3} & \mfrac{1}{4} \\
-\mfrac{1}{4} & \mfrac{1}{4} & \mfrac{2}{3} 
\end{pmatrix}
\quad.
\end{equation}
The eigenvalues of $I_{AB} \omega_B = \lambda \omega_A$ are $\frac{1}{6} M a^2$ and $\frac{11}{12} M a^2$ (twice). The eigenvector for the lowest eigenvalue is $\omega_A = (1,-1,1)$, so the corresponding eigenbivector is
\begin{equation}
\omega_{ij} = \sum_A \omega_A \, (b_A)_{ij} = \begin{pmatrix}
  0 & 1 & -1 \\
  -1 & 0 & 1 \\
  1 & -1 & 0 
\end{pmatrix}
\quad.
\end{equation}
The corresponding pseudovector is $\vec{\omega} = (1,1,1)$.

The bivector formalism also enables us to consider a similar configuration in a four-dimensional space $(x,y,z,w)$: a hypercube with sides of length $a$ along the coordinate axes and a corner at the origin. The density is $\rho = M/a^4$, and we again find $I_{xyxy} = \frac{2}{3} M a^2$, $I_{xyxz} = \frac{1}{4} M a^2$; components like $I_{xyzw}$ with no repeated indices are zero. As discussed in Appendix~\ref{sec:ExtraDims}, in four dimensions there are six independent planes of rotation, and we choose the basis
\begin{equation}
\bivec{b}_A = (\hat{x}\wedge\hat{y},\hat{x}\wedge\hat{z},\hat{y}\wedge\hat{z},\hat{x}\wedge\hat{w},\hat{y}\wedge\hat{w},\hat{z}\wedge\hat{w}) \quad.
\end{equation}
 A full calculation of the inertia tensor in this basis gives 
\begin{equation}
I_{AB} = M a^2 \begin{pmatrix}
\mfrac{2}{3} & \mfrac{1}{4} & -\mfrac{1}{4} & \mfrac{1}{4} & -\mfrac{1}{4} & 0 \\
\mfrac{1}{4} & \mfrac{2}{3} & \mfrac{1}{4} & \mfrac{1}{4} & 0 & -\mfrac{1}{4} \\
-\mfrac{1}{4} & \mfrac{1}{4} & \mfrac{2}{3} & 0 & \mfrac{1}{4} & -\mfrac{1}{4} \\
\mfrac{1}{4} & \mfrac{1}{4} & 0 & \mfrac{2}{3} & \mfrac{1}{4} & \mfrac{1}{4} \\
-\mfrac{1}{4} & 0 & \mfrac{1}{4} & \mfrac{1}{4} & \mfrac{2}{3} & \mfrac{1}{4} \\
0 & -\mfrac{1}{4} & -\mfrac{1}{4} & \mfrac{1}{4} & \mfrac{1}{4} & \mfrac{2}{3}
\end{pmatrix}
\;.
\end{equation}
There are two distinct eigenvalues, $\frac{1}{6} Ma^2$ and $\frac{7}{6} M a^2$, each with multiplicity three.

The eigenspace of the lowest eigenvalue is spanned by bivectors that are equal combinations of three basis elements whose coordinates are in cyclic order: for example, $\omega_A = (1,-1,1,0,0,0)$ corresponds to $\bivec{\omega} = \hat{x}\wedge\hat{y} + \hat{y}\wedge\hat{z} + \hat{z}\wedge\hat{x}$ (precisely the lowest eigenbivector found earlier for the three dimensional case, with the $w$ axis fixed), and $\omega_A = (0,1,0,-1,0,1)$ corresponds to $\bivec{\omega} = \hat{x}\wedge\hat{z} + \hat{z}\wedge\hat{w} + \hat{w}\wedge\hat{x}$.

\subsection{Rotating reference frames}
\label{sec:RotatingFrame}

The equations describing rigid object rotations in Section~\ref{sec:MatrixProduct} is closely related to the ones describing a rotating (non-inertial) reference frame. Newton's second law in a coordinate frame rotating with angular velocity $\bivec{\Omega}$ relative to an inertial frame can be written as a sum of the true force, the Coriolis force, and the centrifugal force:
\begin{equation}
m \ddot{\vec{r}} = \vec{F} + \vec{F}_\text{cor} + \vec{F}_\text{cf} \quad.
\end{equation}
In bivector language, the traditional cross product expression for the Coriolis force $\vec{F}_\text{cor} = 2m\, \dot{\vec{r}} \times \vec{\Omega}$ becomes
\begin{equation}
\vec{F}_\text{cor} = 2m\, \bivec{\Omega} \cdot \dot{\vec{r}}
 = -2m\, \dot{\vec{r}} \cdot \bivec{\Omega}
 \quad.
\end{equation}
As in Section~\ref{sec:MatrixProduct}, the force's direction is found by projecting the velocity to the plane of rotation and then rotating that vector \ang{90} \emph{opposite} the direction of rotation. 

Meanwhile, the traditional expression for the centrifugal force, $\vec{F}_\text{cf} = m \, (\vec{\Omega} \times \vec{r}) \times \vec{\Omega} = 
-m \, \vec{\Omega} \times (\vec{\Omega} \times \vec{r})$ becomes 
\begin{equation}
\vec{F}_\text{cf} = -m\, (\vec{r} \cdot \bivec{\Omega}) \cdot \bivec{\Omega}
 =  -m\, \vec{r} \cdot \bivec{\Omega}^2
 \quad.
\end{equation}
The force's direction exactly matches the projection of the vector to the plane: the dot product projects to the plane, $\bivec{\Omega}{\vphantom{\Omega}}^2$ rotates by \ang{180}, and the minus sign returns to the original projected direction.

On a technical level, these equations are comparable in complexity to the traditional cross product forms. But conceptually, projecting to the plane of rotation (just once!) may be more straightforward to visualize than a series of cross products and right hand rules, and it relates more directly to the rotational motion itself.

\subsection{Quantum commutation relations}
\label{sec:CommutationRelations}

In quantum mechanics, the angular momentum operator $\hat{\vec{J}}$ is the generator of rotations: the rotation operator for angle $\phi$ about (unit vector) axis $\vec{n}$ is $\hat{R}(\vec{n},\phi) = e^{-i\phi \vec{n} \cdot \hat{\vec{J}}/\hbar}$. The essential properties of the angular momentum operator are encoded in its commutation relations: $[\hat{J}_i, \hat{J}_j] = i \hbar \sum_k \epsilon_{ijk} \hat{J}_k$.

All of this can be translated into bivector language in a very straightforward way, with essentially no change other than notation. Angular momentum is a bivector operator $\hat{\bivec{J}}$, and a plane of rotation is specified by a unit bivector $\bivec{n}$. The dot product in the exponent is replaced by a ``double dot product'' $\frac{1}{2}\bivec{n} : \hat{\bivec{J}} = \frac{1}{2} \sum_{i,j=1}^3 n_{ij} \hat{J}_{ij}$ as defined in Eq.~\eqref{eq:pseudo-dot-pseudo-doubledot}. The rotation operator is then
\begin{equation}
\hat{R}(\vec{n},\phi) = e^{-i\phi \bivec{n} : \hat{\bivec{J}}/2\hbar}
\quad.
\end{equation}
The commutation relations become
\begin{align}
  \nonumber
[\hat{J}_{ij}, \hat{J}_{kl}] &= i \hbar \left(
  \delta_{ik} \hat{J}_{jl} - \delta_{il} \hat{J}_{jk} - \delta_{jk} \hat{J}_{il}
    + \delta_{jl} \hat{J}_{ik}
  \right)  \\
\label{eq:JJBivecCommutation}
  &= i \hbar \delta_{ik} \hat{J}_{jl} + (\text{symmetries})
  \quad.
\end{align}
At most one of these four terms will be non-zero: as with Eq.~\eqref{eq:bivector-inertia-tensor}, the four terms just reflect the different ways that an index from the first operator could pair with from the second.

While in three dimensions this expression is more complicated than the vector version, one advantage of this form is that it leads directly to the relativistic generalization. If we allow the indices to run over $\mu,\nu,\ldots = 0,1,2,3$ and replace the $\delta_{ij}$'s with $\eta_{\mu\nu} = \diag(-1,1,1,1)$, these are precisely the commutation relations of the Lorentz group: the fundamental symmetry group of spacetime (see, e.g., Weinberg,\cite{Weinberg:1995mt} Eq.~(2.4.12)). It is very satisfying to see that in some sense, the bivector language for angular momentum already implicitly contains the structure of relativistic boosts and their relationship to rotations.

Finally, because all vectors (and pseudovectors) transform in the same way under rotations, the traditional commutation relations of angular momentum with any vector operator $\hat{\vec{V}}$ match those for $\hat{J}_i$ with itself: $[\hat{J}_i, \hat{V}_j] = i \hbar \epsilon_{ijk} \hat{V}_k$. In terms of the bivector angular momentum, 
\begin{align}
  [\hat{J}_{ij}, \hat{V}_k]
   &= i \hbar \left( \delta_{ik} \hat{V}_j - \delta_{jk} \hat{V}_i \right) 
\label{eq:JVBivecCommutation}
   = i \hbar \delta_{ik} \hat{V}_j - (\text{symmetry}) \;.
\end{align}
As before, the two terms reflect the antisymmetry of $\hat{J}_{ij}$.

The quantum results presented here may not offer many  new insights on their own (though the connection to the Lorentz group is interesting), but we hope they are enough to demonstrate that the building blocks of quantum mechanics are compatible with a bivector formulation.

\section{Sketches of proofs}
\label{app:Proofs}


\subsection{Bivectors as sums of orthogonal tiles}
\label{app:OrthogDecomposition}

Any bivector $\bivec{b}$ in $d$ dimensions can be expressed as a sum of $d/2$ vector wedge products (rounded down) with all of the vectors mutually orthogonal, each corresponding to a tile. Unless two of the tiles have the same magnitude, this orthogonal sum is unique.

A sketch of the argument is as follows.\cite{se:octonion}
The matrix product of $\bivec{b}$ with itself ($\bivec{b}\cdot\bivec{b}$) is a symmetric matrix, so its eigenvectors form a basis. Consider one eigenvector $(\bivec{b}\cdot\bivec{b})\cdot\vec{u} = \lambda \vec{u}$ with $\lambda\ne 0$, scaled for convenience so $|\vec{u}|=1$, and define $\vec{v} \equiv \bivec{b}\cdot\vec{u}$. (This implies that $\vec{v}$ is a second eigenvector with eigenvalue $\lambda$, and since $\bivec{b}$ is antisymmetric, $\vec{u}\cdot\vec{v}=\vec{u}\cdot\bivec{b}\cdot\vec{u}=0$.) 
The antisymmetry of $\bivec{b}$ also implies $\vec{u}\cdot\bivec{b}\cdot\bivec{b}\cdot\vec{u}=-\vec{v}\cdot\vec{v}=-|\vec{v}|^2$, which combines with $\vec{u}\cdot\bivec{b}\cdot\bivec{b}\cdot\vec{u}=\vec{u}\cdot(\lambda\vec{u}) = \lambda$ to show $\lambda = -|\vec{v}|^2$.

Given all this, we can write in components $b_{ij} \equiv v_i u_j - u_i v_j + a_{ij}$, where $\bivec{a}$ is a new bivector that satisfies $\bivec{a}\cdot\vec{u} = \bivec{a}\cdot\vec{v} = 0$ but otherwise has the same eigenvectors and eigenvalues as $\bivec{b}$. Repeat the whole process for $\bivec{a}$, choosing only vectors orthogonal to $\vec{u}$ and $\vec{v}$: in the end, you'll get $\bivec{b} = \vec{v} \wedge \vec{u} + \dotsb$ as desired.
(Eigenvectors with different eigenvalues are orthogonal, so the only way this will fail to be unique is if more than one pair of vectors has the same $\lambda$. Then, \emph{any} decomposition of that eigenspace into orthogonal tiles will work.)

\subsection{Formal addition of tiles}
\label{app:FormalTileAdd}

Section~\ref{sec:BivecAddition} described a procedure for the geometric addition of two simple bivectors $\bivec{a}$ and $\bivec{b}$ (that is, two bivector tiles) in three dimensions. As mentioned there, formally this involves expressing each bivector as a wedge product of vectors, one of which is shared: $\bivec{a} = \vec{u} \wedge \vec{w}$ and $\bivec{b} = \vec{v} \wedge \vec{w}$. Here, we sketch a procedure for finding these vectors that is valid in any dimension $d$.

The first step is to find a unit vector $\vec{w}$ that lies in both bivector planes. Use the eigenvector procedure in Appendix~\ref{app:OrthogDecomposition} to write  $\bivec{a} = \vec{p}_1 \wedge \vec{p}_2$ and  $\bivec{b} = \vec{q}_1 \wedge \vec{q}_2$. Our desired vector $\vec{w}$ lies in both planes, so it must be a linear combination of both pairs of vectors:
\begin{equation}
\label{eq:CommonVectorSystem}
\vec{w} = f_1 \vec{p}_1 + f_2 \vec{p}_2 = g_1 \vec{q}_1 + g_2 \vec{q}_2 \quad.
\end{equation}
This can be interpreted as a system of $d$ equations with unknowns $f_1$, $f_2$, $g_1$, and $g_2$. Specifying $|\vec{w}|=1$ gives one additional equation (with an unimportant $\pm$ sign ambiguity), for a total of $d+1$ equations in four unknowns.

If the two bivectors are coplanar, the system of equations is degenerate: this procedure is not necessary (and any vector in the plane would work). 
Generically in $d>3$ the system is inconsistent: 
the planes are usually linearly independent, so the only solution to Eq.~\eqref{eq:CommonVectorSystem} is the zero vector and the sum cannot be expressed as a single tile. But when $d=3$ or for intersecting planes in higher dimensions, this can be solved for the unit vector $\vec{w}$.

Once $\vec{w}$ is known, the other vectors can be found directly by bivector matrix multiplication as in Section~\ref{sec:MatrixProduct}. Since $\vec{w}$ lies in the plane of each tile, there is no projection step: $\vec{u} = \bivec{a} \cdot \vec{w}$ is rotated \ang{90} from $\vec{w}$ in the plane of $\bivec{a}$, and it has magnitude $|\vec{u}| = |\bivec{a}|$ because $|\vec{w}|=1$. Similarly, $\vec{v} = \bivec{b} \cdot \vec{w}$.

With these vectors all established, we can finally write $\bivec{a} + \bivec{b} = \vec{u} \wedge \vec{w} + \vec{v} \wedge \vec{w} = (\vec{u}+\vec{v}) \wedge \vec{w}$, each step of which can be visualized geometrically as in Figure~\ref{fig:tileAdd}.

\section{Bivector equivalents of pseudovector products}
\label{app:BivecRelations}

We collect here a number of formal results relating pseudovector product equations to their bivector equivalents. The results are presented in index notation and involve the totally antisymmetric Levi-Civita symbol $\epsilon_{ijk}$, where $\epsilon_{xyz}=+1$, $\epsilon_{yxz}=-1$, etc.
As we will see, the final bivector forms do not involve $\epsilon_{ijk}$ at all. Although the derivations here are entirely in Cartesian three dimensional space for clarity, the final bivector results in index notation generalize directly to curved space in any dimension (where bivectors are antisymmetric rank-2 contravariant tensors).

For clarity, in this appendix we will use the symbols $\vPv$, $\vQv$, and $\vRv$ to refer to generic ordinary vectors (``polar vectors''), while the symbols $\pvBv$, $\pvCv$, and $\pvDv$ will refer to generic pseudovectors (``axial vectors'') whose bivector equivalents are $\bvBv$, $\bvCv$, and $\bvDv$, respectively. In most cases, the equivalences can be derived using the identity
\begin{equation}
\label{eq:EpsilonSquared}
\sum_{i=1}^3 \epsilon_{ijk} \epsilon_{i\idxm\idxn} = \delta_{j\idxm} \delta_{k\idxn} - \delta_{j\idxn} \delta_{k\idxm} \quad,
\end{equation}
though at times a more general identity is needed:
\begin{align}
\label{eq:GenEpsilonSquared}
\nonumber
\epsilon_{ijk} \epsilon_{\idxl\idxm\idxn}
 = {}&\delta_{i\idxl} \delta_{j\idxm} \delta_{k\idxn}
  + \delta_{i\idxm} \delta_{j\idxn} \delta_{k\idxl}
  + \delta_{i\idxn} \delta_{j\idxl} \delta_{k\idxm} \\
  & - \delta_{i\idxl} \delta_{j\idxn} \delta_{k\idxm}
  - \delta_{i\idxm} \delta_{j\idxl} \delta_{k\idxn}
  - \delta_{i\idxn} \delta_{j\idxm} \delta_{k\idxl} \;.
\end{align}

For any pseudovector $\pvBv$ (in three dimensions), we can define a unique ``dual'' bivector $\bvBv$ whose plane is normal to $\pvBv$ , whose magnitude (``area'': $|\bvBv|$) is equal to the vector's magnitude (``length'': $|\pvBv|$), and whose orientation relates to the direction of $\pvBv$ by the right hand rule. The matrix components $\bvBl_{ij}$ that describe this dual bivector are:
\begin{equation}
\label{eq:vec-to-bivec}
\bvBl_{ij} = \sum_{k=1}^3 \epsilon_{ijk} \pvBl_k
 \equiv \epsilon_{ijk} \pvBl_k \quad.
\end{equation}
This second form illustrates the use of Einstein summation notation (used implicitly through the rest of this section) where repeated indices within a single term are assumed to be summed over all coordinates.%
\footnote{For simplicity, throughout this work we assume Cartesian orthonormal coordinates, so the distinction between covariant and contravariant indices can be ignored.} 
Explicitly in terms of components, this reads
\begin{equation}
\label{eq:bivec-vec-components}
\bvBl_{ij} 
 = \begin{pmatrix}
0 & \bvBl_{xy} & -\bvBl_{zx} \\
-\bvBl_{xy} & 0 & \bvBl_{yz} \\
\bvBl_{zx} & -\bvBl_{yz} & 0 
\end{pmatrix} = \begin{pmatrix}
0 & \pvBl_z & -\pvBl_y \\
-\pvBl_z & 0 & \pvBl_x \\
\pvBl_y & -\pvBl_x & 0 
\end{pmatrix} \quad.
\end{equation}
We can see that $\bvBl_{ji} = -\bvBl_{ij}$, because of the antisymmetry of $\epsilon_{ijk}$. Comparing components between the two matrices makes the geometric significance clear: for example, the bivector component $\bvBl_{xy}$ in the $xy$-plane is perpendicular to the vector component $\pvBl_z$, with the same magnitude.

The relationship also works in the other direction, with a factor of a half to compensate for double counting (since each independent bivector component appears twice in the matrix):
\begin{equation}
\label{eq:vec-bivec-components}
\pvBl_i = \frac{1}{2} 
         \epsilon_{ijk} \bvBl_{jk}
 = \begin{pmatrix} \bvBl_{yz} \\ \bvBl_{zx} \\ \bvBl_{xy} \end{pmatrix}
 \quad.
\end{equation}

\begin{figure}
\centering
\begin{overpic}[percent,width=0.8\linewidth]{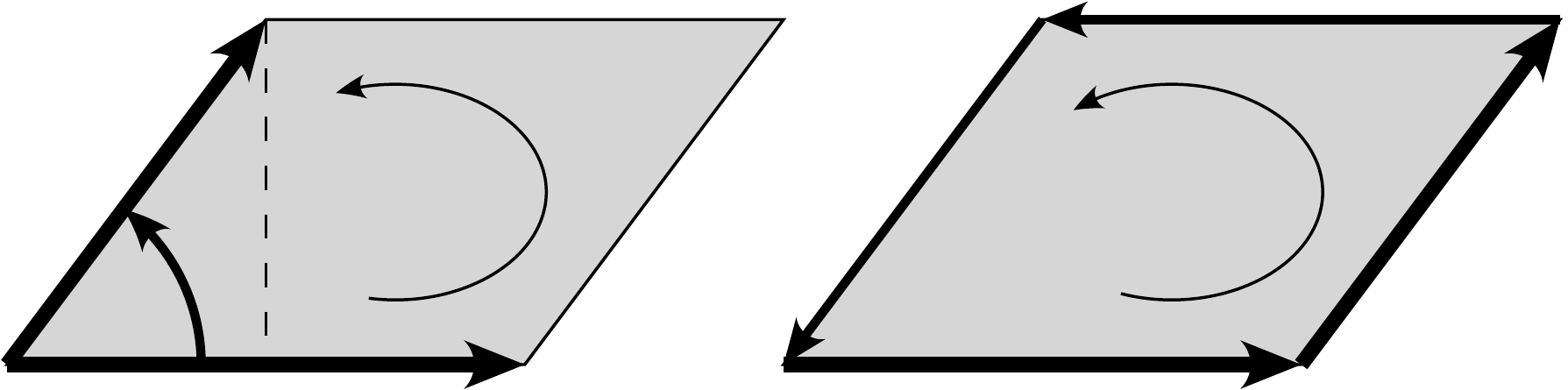}%
  \put(18,-4){$\vPv$}
  \put(7,16){$\vQv$}
  \put(7,4){$\theta$}
  \put(19,11){$h$} 
  \put(65,-4){$\vPv$}
  \put(93,9){$\vQv$}
\end{overpic}
\caption{\label{fig:WedgeTiles} In the wedge product $\bvBv = \vPv \wedge \vQv$, the two vectors define a parallelogram of area $|\bvBv| = |\vPv| h = |\vPv| |\vQv| \sin\theta$ with a fixed attitude in space. This figure illustrates two ways to visualize its orientation: either as the direction that rotates the first vector toward the second tail-to-tail, or by following the first vector tip-to-tail with the second around the edges.}
\end{figure}
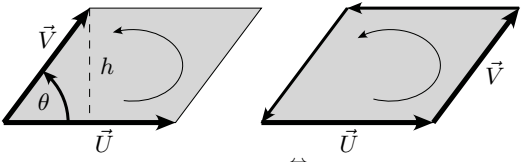

We begin with cross products, since that is where pseudovectors arise. 
The cross product of two ordinary vectors $\pvBv = \vPv \times \vQv$ is expressed in index notation as
\begin{equation}
\pvBl_i = 
 \epsilon_{ijk} \vPl_j \vQl_k \quad.
\end{equation}
Using the definition in Eq.~\eqref{eq:vec-to-bivec} (after suitably relabeling summed indices), the symmetries of $\epsilon_{ijk}$, and the identity \eqref{eq:EpsilonSquared}, we find
\begin{align}
\bvBl_{ij} &= 
               \epsilon_{ijk} \pvBl_k
 = 
     \epsilon_{ijk}
    \left( 
            \epsilon_{k\idxm\idxn} \vPl_\idxm \vQl_\idxn \right)
 = 
    \left( 
            \epsilon_{kij} \epsilon_{k\idxm\idxn} \right)
     \vPl_\idxm \vQl_\idxn
\nonumber \\
\label{eq:vec-cross-vec-indices}
&= 
    \left( \delta_{i\idxm} \delta_{j\idxn} - \delta_{i\idxn} \delta_{j\idxm} \right)
    \vPl_\idxm \vQl_\idxn
 = \vPl_i \vQl_j - \vPl_j \vQl_i \quad.
\end{align}
This is the definition of the wedge product:
\begin{equation}
\label{eq:vec-cross-vec-wedge}
\boxed{%
\pvBv = \vPv \times \vQv \qquad \longleftrightarrow \qquad
\bvBv = \vPv \wedge \vQv%
}
\quad.
\end{equation}
This can be visualized as explained in Figure~\ref{fig:WedgeTiles}.

\begin{figure}
\centering
\begin{overpic}[percent,width=3.5cm]{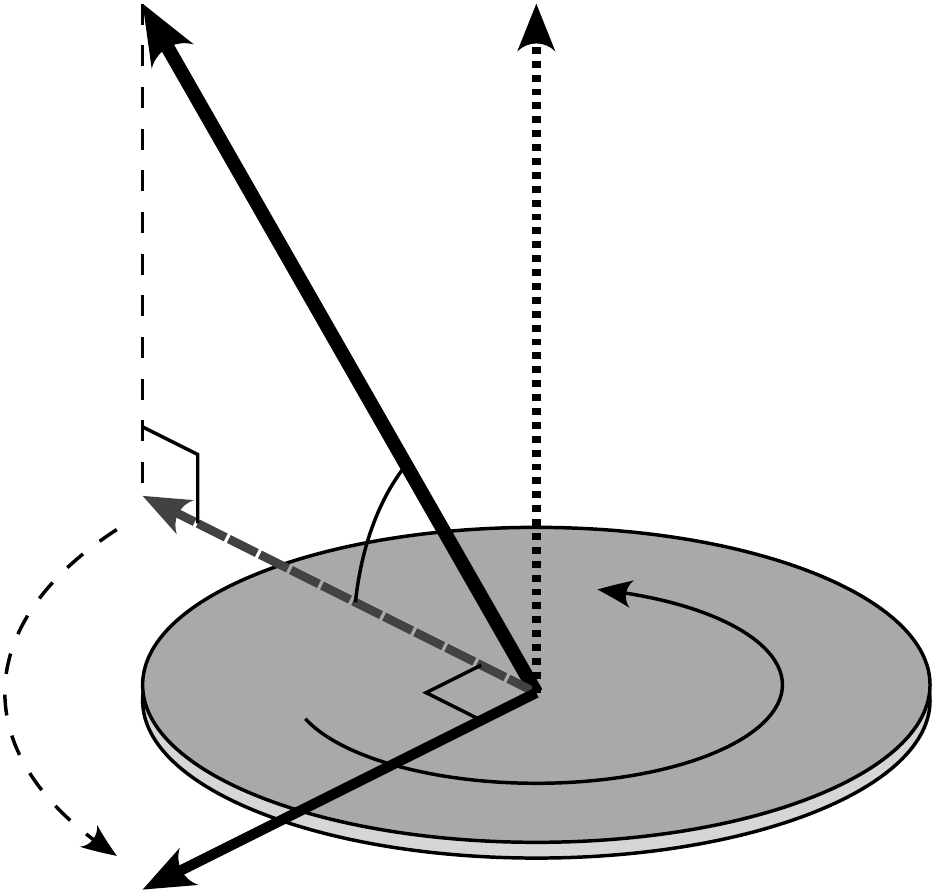}
  \put(67,18){$\bvBv$}
  \put(60,82){$\pvBv$}
  \put(30,79){$\vPv$}
  \put(21,-7){$\vPv \cdot \bvBv$}
  \put(33,40){$\theta$}
\end{overpic}
\vspace{1mm}
\caption{\label{fig:VecTileMult} Finding the (dot) product of a vector $\vPv$ with a bivector $\bvBv$ is a two-step process. First, project the vector into the plane of the bivector, and then rotate it \ang{90} in the tile's orientation direction. The resulting magnitude is $|\vPv|\,|\bvBv|\,\cos\theta$, where $\theta$ is the angle between the vector and the bivector plane, and the $\cos\theta$ comes from the projection. (This matches the formula for cross product magnitude: if $\pvBv$ is the pseudovector corresponding to $\bvBv$, then its angle from $\vPv$ is $\ang{90}-\theta$ and thus $|\pvBv\times\vPv|=|\pvBv|\,|\vPv|\,\sin(\ang{90} - \theta)=|\pvBv|\,|\vPv|\,\cos\theta$.)}
\end{figure}

Meanwhile, the cross product of  a pseudovector and an ordinary vector is an ordinary vector, $\vQv = \pvBv \times \vPv$, and can be expressed in index notation as
\begin{equation}
\label{eq:pseudo-cross-vec-indices}
\vQl_i
 = 
  \epsilon_{ijk} \pvBl_j \vPl_k = 
    \vPl_k \left(
    \epsilon_{kij} \pvBl_j \right)
  = 
  \vPl_k \bvBl_{ki} \quad.
\end{equation}
(The most familiar physical example of this is the magnetic force on a charged particle: $\vec{F} = q \vec{v} \times \vec{B}$, where the magnetic field $\vec{B}$ is a pseudovector.)
We can immediately recognize this as matrix multiplication of a row vector times a square matrix, which we will represent symbolically as a dot product:
\begin{equation}
\label{eq:vec-cross-pseudo-matrix}
\boxed{%
\vQv =  \pvBv \times \vPv \qquad \longleftrightarrow \qquad
\vQv = \vPv \cdot \bvBv%
}
\quad.
\end{equation}
This can be visualized as explained in Figure~\ref{fig:VecTileMult}. (Note that the order of the two terms has reversed.)
Given the antisymmetry of $\bvBv$, if we reinterpret $\vPv$ as a column vector we can instead write the matrix product in the opposite order: $\vQv = -\bvBv \cdot \vPv$. Performing the matrix multiplication often feels more familiar in this order.

\textit{(In the remainder of this section, we will omit the derivations of each bivector expression: the methods are similar to those above, but the details can get tedious.)}

\begin{figure}
\centering
\begin{overpic}[percent,width=\linewidth]{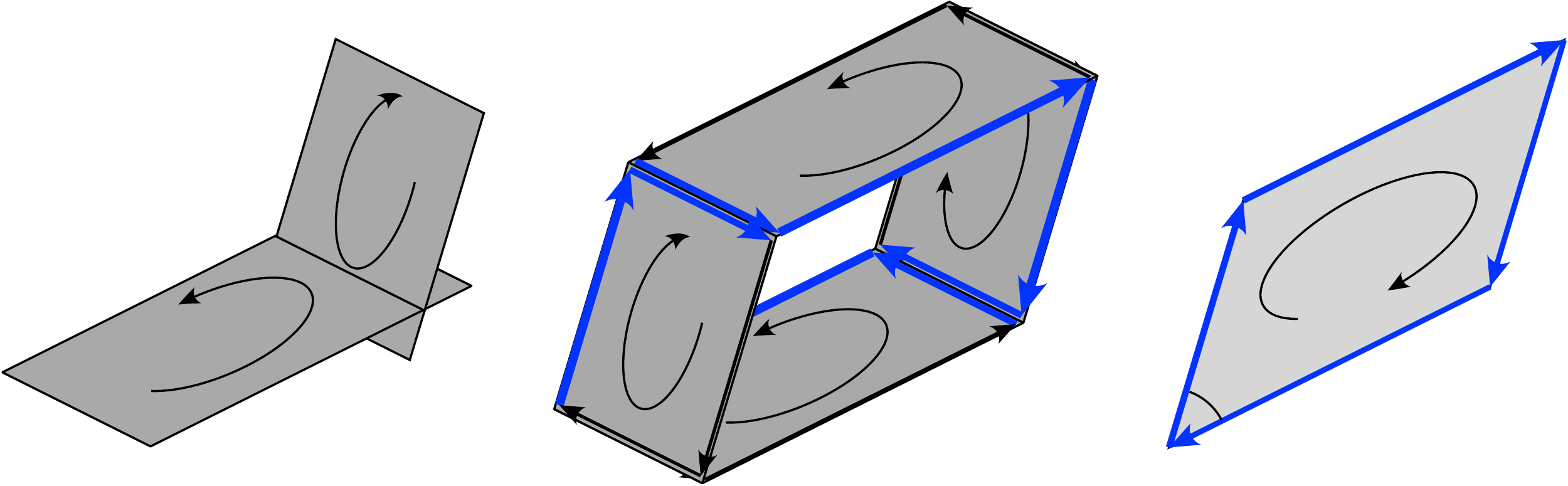}
  \put(15,0){$\bvBv$}
  \put(15,22){$\bvCv$}
  \put(86,3){$[\bvCv,\bvBv]$}
  \put(78,6.5){$\theta$}
\end{overpic}
\caption{\label{fig:TileCommute} The commutator $[\bvCv,\bvBv]$ of two tiles is non-zero if their planes intersect along a line (with angle $\theta$ between them). Reshape the tiles as rectangles with unit length along the shared line, and form a ``pipe'' out of two copies of each as shown. The cross section of the pipe (e.g.\ either ``end cap'') is the tile representing the commutator: a parallelogram of area $|\bvCv|\,|\bvBv| \sin\theta$. As for orientation, find an edge of the ``pipe'' where the overlapping edges have the same orientation: the orientation of the result matches the edge from the first tile ($\bvCv$) that points \emph{into} the shared edge and the edge of the second ($\bvBv$) that points \emph{out} of it. Imagining that shared edges with opposite orientation ``cancel out,'' you can follow the arrows around the pipe for a full cycle. (Reversing the order of the commutator changes which path around the pipe you follow.)}
\end{figure}
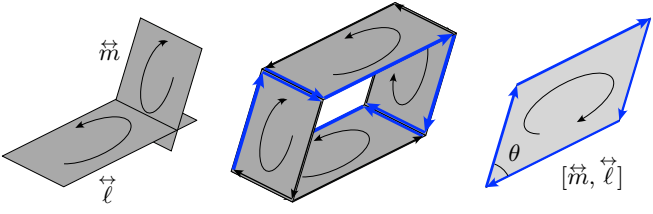

Although it is less common in physical formulas, for completeness we should also consider the cross product of two pseudovectors, whose result is another pseudovector: $\pvDv = \pvBv \times \pvCv$. Converting this to bivector language is messy: it starts like Eq.~\eqref{eq:vec-cross-vec-indices} and uses Eqs.~\eqref{eq:vec-bivec-components} and~\eqref{eq:GenEpsilonSquared}, with the result
\begin{equation}
\bvDl_{ij} =  
  \left( \bvCl_{ik} \bvBl_{kj} - \bvBl_{ik} \bvCl_{kj} \right) \quad.
\end{equation}
The result is recognizable as the matrix product of the two bivectors, antisymmetrized:
\begin{equation}
\label{eq:pseudo-cross-pseudo-commutator}
\boxed{%
\pvDv = \pvBv \times \pvCv \quad \longleftrightarrow \quad
\bvDv = \bvCv \cdot \bvBv - \bvBv \cdot \bvCv
  \equiv [\bvCv,\bvBv]%
}
\quad.
\end{equation}
The final square bracket notation reflects the fact that this is the commutator of the two matrices. For ``simple'' bivectors that can each be represented by a single tile, this can be visualized as described in Figure~\ref{fig:TileCommute}.

\medskip

Having established the formulas for cross products involving pseudovectors, we next consider dot products. The dot product of two pseudovectors $\alpha = \pvBv \cdot \pvCv$ is a scalar and can be written as
\begin{equation}
\label{eq:pseudo-dot-pseudo-components}
\alpha = 
  \pvBl_i \pvCl_i = \tfrac{1}{2} 
  \bvBl_{jk} \bvCl_{jk} \quad.
\end{equation}
As in previous equations, the factor of $\frac{1}{2}$ compensates for the duplication of entries in the two antisymmetric matrices. This sum over products of all matrix components is sometimes called the ``double dot product'' and is denoted by a colon:%
\footnote{In this work, the (single) dot product always represents matrix multiplication, or more precisely, tensor index contraction on one index. The double dot product notation used here for contraction on two indices is more common in engineering than in physics: in the traditional physics curriculum, students see second rank tensors only rarely before they learn index notation, at which point these shorthands become unnecessary. In geometric algebra the convention is different: there, roughly speaking, a (single) dot product denotes contraction on as many indices as possible, so the double index contraction of two bivectors would be represented by a single dot product.}
\begin{equation}
\label{eq:pseudo-dot-pseudo-doubledot}
\boxed{%
\alpha = \pvBv \cdot \pvCv \qquad \longleftrightarrow \qquad
\alpha = \tfrac{1}{2} \bvBv \dblcdot \bvCv%
}
\quad.
\end{equation}
As with the vector dot product, we can interpret this as a measure of whether two bivectors are oriented in the same direction: orthogonal tiles have double dot product zero.
We can use this to define the magnitude of a bivector: $|\bivec{\bvBl}|^2 \equiv \frac{1}{2} \sum_{i,j=1}^3 \bvBl_{ij}^2$.
If we use the antisymmetry of the bivector matrices, we can also interpret this in terms of the trace of the matrix product: $\alpha = -\frac{1}{2} \tr(\bvBv \cdot \bvCv)$.
And if we express $\bvCv = \vPv \wedge \vQv$, then $\alpha = \vPv \cdot \bvBv \cdot \vQv$.

Geometrically, this is a measure of the degree to which the two planes have the same attitude and orientation in space. For two tiles whose planes overlap along a line (which includes any two tiles in three dimensions) the double dot product matches the dot product of their normal vectors: $\frac{1}{2} \bvBv \dblcdot \bvCv = |\bvBv|\, | \bvCv| \cos\theta$, where $\theta$ is the angle between the planes. More generally in higher dimensions (where two planes may only overlap at a point), the double dot product equals zero if \emph{any} vector in one tile is perpendicular to the other tile. The precise geometric meaning in general seems complicated: considering the form $\vPv \cdot \bvBv \cdot \vQv$ in light of Figure~\ref{fig:VecTileMult}, we are projecting $\vPv$ into the plane of $\bvBv$, multiplying it by the bivector magnitude, rotating it \ang{90}, and then taking the dot product with $\vQv$.

Finally, the dot product of an ordinary vector and a pseudovector $\phi = \vPv \cdot \pvBv$ is more subtle to understand, because the result $\phi$ is a ``pseudoscalar'' rather than an ordinary scalar:
\begin{align}
\phi &= 
 \vPl_i \pvBl_i
 = \tfrac{1}{2} 
 \epsilon_{ijk} \vPl_i \bvBl_{jk}
 \nonumber \\
 \label{eq:vec-dot-pseudo}
&=  \vPl_x \bvBl_{yz} + \vPl_y \bvBl_{zx} + \vPl_z \bvBl_{xy} \quad.
\end{align}
Every term contains all three spatial coordinates: there are no free indices, just like a scalar quantity. But like a pseudovector, under a reflection like $x \to -x$ all three terms change sign, which a true scalar would not.

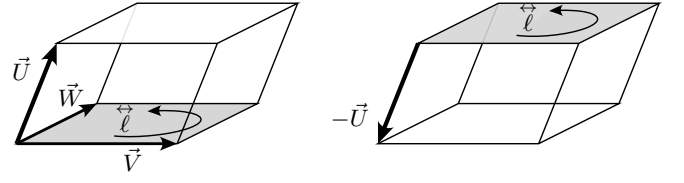
\begin{figure}
\centering
\begin{overpic}[percent,width=\linewidth]{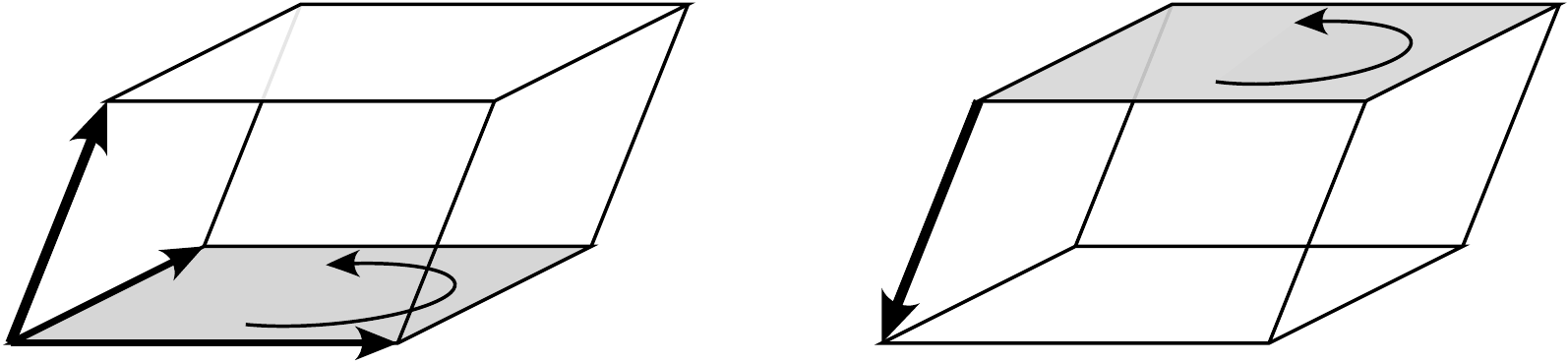}
  \put(16,3.){$\bvBv$}
  \put(0,11){$\vPv$}
  \put(17,-3){$\vQv$}
  \put(7,7){$\vRv$}
  \put(78,18.5){$\bvBv$}
  \put(49,4){$-\vPv$}
\end{overpic}
\vspace{-0.75em}
\caption{\label{fig:Trivector} The wedge product of a vector with a bivector, $\vPv \wedge \bvBv$ (shown at left), can be visualized as an oriented three-dimensional volume formed by extruding the bivector tile along the vector's direction. If the bivector is a wedge product $\bvBv = \vQv \wedge \vRv$, this defines the triple product $\vPv \wedge \vQv \wedge \vRv$. If we choose right-handed coordinates, the component $\phi \equiv \Phi_{xyz}$ is positive if the bivector tile's orientation appears counterclockwise from inside the region (as shown at left) and negative if it appears clockwise from inside (as with $-\vPv \wedge \bvBv$ at right).}
\end{figure}
The appropriate interpretation (which correctly generalizes to higher dimensions) is not a scalar, but the wedge product of $\vPv$ with $\bvBv$:
\begin{subequations}
\label{eq:vec-dot-pseudo-trivec}
\begin{align}
\boxed{%
\phi = \vPv \cdot \pvBv \qquad \longleftrightarrow \qquad
\Phi_{(3)} = \vPv \wedge \bvBv } \\
\label{eq:TrivecIndices}
\Phi_{ijk} = \vPl_i \bvBl_{jk} + \vPl_j \bvBl_{ki} + \vPl_k \bvBl_{ij}%
\quad.
\end{align}
\end{subequations}
This is a \defn{trivector}: a totally antisymmetric rank-3 tensor. (The product is symmetric: $\vPv \wedge \bvBv = \bvBv \wedge \vPv$.) As shown in Figure~\ref{fig:Trivector}, this can be visualized as an (oriented) region of 3D space, with the bivector tile $\bvBv$ as its base and the vector $\vPv$ showing how that base ``extrudes'' into the third dimension. 

If the bivector is a wedge product $ \bvBv = \vQv \wedge \vRv$, this defines the ``triple product'' of (true) vectors: $\vPv \wedge \vQv \wedge \vRv$.
By convention, the pseudoscalar is defined to equal the specific component $\phi \equiv \Phi_{xyz}$ of the trivector: the coefficient of $\hat{x}\wedge\hat{y}\wedge\hat{z}$. (In a right-handed coordinate system, the sign of $\phi$ is positive if the bivector's orientation looks counterclockwise when viewed from inside the parallelepiped.)

Because Equation~\eqref{eq:TrivecIndices} is totally antisymmetric in $i$, $j$, and $k$, it is zero unless the three indices are different. In three dimensions, this means that Eq.~\eqref{eq:vec-dot-pseudo} is the only independent term: that's why it looks like a scalar. But in $d$ dimensions, there are $\binom{d}{3} = \frac{1}{3!} d (d-1)(d-2)$ independent ways of choosing three coordinate labels out of $d$, so that is the number of components. 


\begin{thebibliography}{10}

\bibitem{Petrov:1954}
A.~Z. Petrov, ``Klassifikacya prostranstv opredelyayushchikh polya
  tyagoteniya,'' {\em Uch. Zapiski Kazan. Gos. Univ.} {\bfseries 114} no.~8,
  (1954) 55--69.

\bibitem{Petrov:2000bs}
A.~Z. Petrov, ``The classification of spaces defining gravitational fields,''
  \href{http://dx.doi.org/10.1023/A:1001910908054}{{\em Gen. Rel. Grav.}
  {\bfseries 32} no.~8, (2000) 1665--1685}. ({E}nglish translation of
  {P}etrov's 1954 paper.).

\bibitem{Hestenes:2003}
D.~Hestenes, ``Oersted medal lecture 2002: {R}eforming the mathematical
  language of physics,'' \href{http://dx.doi.org/10.1119/1.1522700}{{\em
  American Journal of Physics} {\bfseries 71} no.~2, (2003) 104--121}.

\bibitem{Gull:1993}
S.~F. Gull, A.~N. Lasenby, and C.~J.~L. Doran., ``Imaginary numbers are not
  real - the geometric algebra of spacetime,'' {\em Found. Phys.} {\bfseries
  23} no.~9, (1993) 1175--1201.

\bibitem{Doran:2003}
C.~Doran and A.~Lasenby, {\em Geometric Algebra for Physicists}.
\newblock Cambridge University Press, Cambridge, 2003.

\bibitem{Trautman:2006fp}
A.~Trautman, ``{E}instein-{C}artan theory,'' in {\em Encyclopedia of
  Mathematical Physics}, J.-P. Francoise, G.~Naber, and {Tsou S.T.}, eds.,
  vol.~2, pp.~189--195.
\newblock Elsevier, 6, 2006.
\newblock \href{http://arxiv.org/abs/gr-qc/0606062}{{\ttfamily
  arXiv:gr-qc/0606062}}.

\bibitem{Deprez:2019}
T.~Deprez, S.~E. Gijsen, J.~Deprez, and M.~De~Cock, ``Investigating student
  understanding of cross products in a mathematical and two electromagnetism
  contexts,''
  \href{http://dx.doi.org/10.1103/PhysRevPhysEducRes.15.020132}{{\em Phys. Rev.
  Phys. Educ. Res.} {\bfseries 15} (Sep, 2019) 020132}.

\bibitem{Kustusch:2016}
M.~B. Kustusch, ``Assessing the impact of representational and contextual
  problem features on student use of right-hand rules,''
  \href{http://dx.doi.org/10.1103/PhysRevPhysEducRes.12.010102}{{\em Phys. Rev.
  Phys. Educ. Res.} {\bfseries 12} (Jan, 2016) 010102},
  \href{http://arxiv.org/abs/1507.02364}{{\ttfamily arXiv:1507.02364}}.

\bibitem{Mashood_2012}
K.~K. Mashood and V.~A. Singh, ``An inventory on rotational kinematics of a
  particle: unravelling misconceptions and pitfalls in reasoning,''
  \href{http://dx.doi.org/10.1088/0143-0807/33/5/1301}{{\em European Journal of
  Physics} {\bfseries 33} no.~5, (Jul, 2012) 1301--1312}.

\bibitem{Note1}
Students accustomed to working with bivector tiles also have an especially easy
  time understanding the relationship between Kepler's second law and angular
  momentum, since the triangle formed by $\protect \vec {r}$ and $\protect \vec
  {v}\protect \,dt$ is essentially just half of the $\protect \bivec {\ell } =
  \protect \vec {r} \wedge \protect \vec {p}$ parallelogram.

\bibitem{Vold:1993rot}
T.~G. Vold, ``An introduction to geometric algebra with an application in rigid
  body mechanics,'' \href{http://dx.doi.org/10.1119/1.17201}{{\em American
  Journal of Physics} {\bfseries 61} no.~6, (1993) 491--504}.

\bibitem{Vold:1993EM}
T.~G. Vold, ``An introduction to geometric calculus and its application to
  electrodynamics,'' \href{http://dx.doi.org/10.1119/1.17202}{{\em American
  Journal of Physics} {\bfseries 61} no.~6, (1993) 505--513}.

\bibitem{Note2}
As of this writing, the Wikipedia entry for bivectors (\protect \url
  {https://en.wikipedia.org/wiki/Bivector}) is entirely about geometric
  algebra. Similarly, the website \protect \texttt {\protect \href
  {https://bivector.net/}{bivector.net}} is actually a geometric algebra site.

\bibitem{Note3}
For a bivector defined in terms of the wedge product introduced in this
  section, it is straightforward to illustrate that changing the
  parallelogram's shape without changing its area leaves the result unchanged.
  For example, because the wedge product is linear in each argument, the
  bivector quantity $6 (\protect \hat {x} \wedge \protect \hat {y})$ can be
  written as $(6\protect \hat {x}) \wedge \protect \hat {y}$, $(2\protect \hat
  {x}) \wedge (3\protect \hat {y})$, $(3\protect \hat {x}-\protect \hat {y})
  \wedge (2\protect \hat {y})$, or infinitely many other forms, each
  corresponding to a differently shaped parallelogram with the same area.

\bibitem{Conant:1974gp}
D.~R. Conant and W.~A. Beyer, ``Generalized pythagorean theorem,''
  \href{http://dx.doi.org/10.2307/2319528}{{\em The American Mathematical
  Monthly} {\bfseries 81} no.~3, (1974) 262--265}.

\bibitem{Note4}
Intuitively, the idea behind this projection-rotation process is that the
  bivector is an area spanned by one vector sweeping around to a second vector.
  By taking a dot product with an additional vector from the left, we replace
  the first vector with a scalar, rotating the result toward the second. This
  can be made formal by reshaping the tile as a wedge product of a unit vector
  along the additional vector's projection and a second vector perpendicular to
  it. (A dot product from the right would rotate the resulting vector \ang {90}
  \protect \emph {opposite} the bivector's orientation, reversing the sign.).

\bibitem{Note5}
Related to this, a rotation matrix can be constructed as the matrix exponential
  $\protect \mathsf {R}(t) = e^{-\protect \bivec {\omega } t}$. Then for a
  point with initial position $\protect \vec {r}_0$, the matrix product
  $\protect \vec {r}(t) = \protect \mathsf {R}(t) \cdot \protect \vec {r}_0$
  gives the rotated position as a function of time.

\bibitem{Penrose:2004rr}
R.~Penrose, {\em The Road to Reality}.
\newblock Random House, New York, 2004.

\bibitem{Note6}
One way of understanding this is that velocities in relativity do not add in a
  linear way: the Einstein velocity transformation means that different atoms
  of a rotating object will have their velocities changed by different amounts
  when boosted to a different frame, so Eq.~\protect \eqref
  {eq:rotationVelocityBivec} will no longer be true (even relative to the
  center of mass).

\bibitem{Jensen:2023mag}
S.~Jensen, ``Teaching magnetism with bivectors.'' (in preparation).

\bibitem{Note7}
If we do not demand that the two planes be mutually orthogonal, there are many
  other ways to write this final angular velocity as a sum of two wedge
  products. For example, $\protect \bigl [ (2\protect \hat {x}+\protect \hat
  {z})\wedge (\protect \hat {y}+2\protect \hat {z}) + (2\protect \hat
  {y}-\protect \hat {z})\wedge (\protect \hat {y}-\protect \hat {w} ) \protect
  \bigr ] \unit {\radian \per \second }$.

\bibitem{minutephysics:flat}
H.~Reich, ``Why is the solar system flat?''
\newblock \url{https://www.youtube.com/watch?v=tmNXKqeUtJM}. {MinutePhysics}.

\bibitem{Note8}
Index notation provides an efficient shorthand for this result: $I_{ijkl} = 4
  \intop dm \protect \, x_{[j} \delta _{i][k} x_{l]}$, where square brackets
  denote antisymmetrization: $M_{[ij]}=\protect \frac {1}{2} (M_{ij} -
  M_{ji})$.

\bibitem{Note9}
Curiously, these are exactly the algebraic symmetries of the Riemann curvature
  tensor.

\bibitem{Note10}
This ``alphabetical order'' basis is one natural choice. If instead we chose
  the ``cyclic'' basis $(\protect \hat {y}\wedge \protect \hat {z}, \protect
  \hat {z}\wedge \protect \hat {x}, \protect \hat {x}\wedge \protect \hat
  {y})$, the corresponding pseudovector basis would be the traditional
  $(\protect \hat {x}, \protect \hat {y}, \protect \hat {z})$ and the
  components of $I_{AB}$ would be the same as in a traditional calculation.

\bibitem{Weinberg:1995mt}
S.~Weinberg, {\em The Quantum theory of fields. Vol. 1: Foundations}.
\newblock Cambridge University Press, 6, 2005.

\bibitem{se:octonion}
{Stack Exchange user `octonion' (users/181126/octonion)}, ``Bivector into
  orthogonal components.'' Mathematics stack exchange.
\newblock \url{https://math.stackexchange.com/q/1039217}.

\bibitem{Note11}
For simplicity, throughout this work we assume Cartesian orthonormal
  coordinates, so the distinction between covariant and contravariant indices
  can be ignored.

\bibitem{Note12}
In this work, the (single) dot product always represents matrix multiplication,
  or more precisely, tensor index contraction on one index. The double dot
  product notation used here for contraction on two indices is more common in
  engineering than in physics: in the traditional physics curriculum, students
  see second rank tensors only rarely before they learn index notation, at
  which point these shorthands become unnecessary. In geometric algebra the
  convention is different: there, roughly speaking, a (single) dot product
  denotes contraction on as many indices as possible, so the double index
  contraction of two bivectors would be represented by a single dot product.

\end{thebibliography}
\begingroup
\endgroup

\end{document}